\documentclass[11pt,a4paper,twoside]{article}
\usepackage[T1]{fontenc}
\usepackage[ansinew]{inputenc}
\usepackage[italian,english]{babel}
\usepackage{amsfonts}
\usepackage{amsmath}
\usepackage{array}
\usepackage{amsthm}
\usepackage{amssymb}
\usepackage{graphicx}
\usepackage{braket}
\usepackage{eucal}
\usepackage{verbatim}
\usepackage{multicol}
\usepackage{tikz}
\usetikzlibrary{arrows,shapes}
\usetikzlibrary{matrix,arrows}
\newcommand{\be}{\begin{eqnarray}}
\newcommand{\ee}{\end{eqnarray}}
\newcommand{\pro}[2]{\mbox{$\langle\, #1 \mid #2\,\rangle$}}
\newcommand{\expec}[1]{\mbox{$\langle\, #1\,\rangle$}}

\renewcommand{\d}{\mbox{${\rm d}$}} %d differenziale non corsivo in math mode
\newcommand{\lp}{\ell_{\rm p}}
\newcommand{\mpl}{m_{\rm p}}
\newcommand{\gn}{G_{\rm N}}
\newcommand{\rh}{r_{\rm H}}
\newcommand{\Rh}{R_{\rm H}}
\title{\bf Black holes as self-sustained quantum states, and Hawking radiation}
\author{Roberto~Casadio$^{ab}$\thanks{E-mail: casadio@bo.infn.it},
$\ $
Andrea~Giugno$^{ab}$\thanks{E-mail: andrea.giugno2@unibo.it},
$\ $
Octavian~Micu$^{c}$\thanks{E-mail: octavian.micu@spacescience.ro},
$\ $
and
Alessio~Orlandi$^{ab}$\thanks{E-mail: orlandi@bo.infn.it}
\\
\\
{\em $^a$Dipartimento di Fisica e Astronomia, Universit\`a di Bologna}
\\
{\em via Irnerio~46, I-40126 Bologna, Italy}
\\
\\
{\em $^b$I.N.F.N., Sezione di Bologna,}
\\
{\em via B.~Pichat~6/2, I-40127 Bologna, Italy}
\\
\\
{\em $^c$Institute of Space Science, Bucharest,}
\\
{\em P.O.~Box MG-23, RO-077125 Bucharest-Magurele, Romania}
}
\begin{document}
\maketitle
\begin{abstract}
We employ the recently proposed formalism of the ``horizon wave-function''
to investigate the emergence of a horizon in models of black holes as
Bose-Einstein condensates of gravitons.
We start from the Klein-Gordon equation for a massless scalar
(toy graviton) field coupled to a static matter current.
The (spherically symmetric) classical field reproduces the Newtonian potential
generated by the matter source, and the corresponding quantum state is given
by a coherent superposition of scalar modes with continuous occupation number.
Assuming an attractive self-interaction that allows for bound states,
one finds that (approximately) only one mode is allowed, and the system can
be confined in a region of the size of the Schwarzschild radius.
This radius is then shown to correspond to a proper horizon,
by means of the horizon wave-function of the quantum system,
with an uncertainty in size naturally related to the expected typical
energy of Hawking modes.
In particular, this uncertainty decreases for larger black hole mass
(with larger number of light scalar quanta), 
in agreement with semiclassical expectations, a result which does
not hold for a single very massive particle.
We finally speculate that a phase transition should occur during the
gravitational collapse of a star, ideally represented by a static matter
current and Newtonian potential, that leads to a black hole, again ideally
represented by the condensate of toy gravitons, and suggest
an effective order parameter that could be used to investigate this transition. 
%\par
%\null
%\par
%\textit{PACS - ...}
\end{abstract}
\section{Introduction}
\setcounter{equation}{0}
\label{intro}
Recent works by Dvali and Gomez have offered a new perspective
on the quantum aspects of black hole physics~\cite{DvaliGomez},
and are drawing more and more
attention~\cite{Kuhnel:2014oja, kuhnel2, Mueck:2013wba,
Casadio:2013hja, Berkhahn:2013woa, flassig, Binetruy:2012kx,mueckPT}.
The idea is very simple:
a black hole can be modelled as a Bose-Einstein
condensate (BEC) of gravitons interacting with each other.
Such gravitons superpose in a single small region of space,
effectively giving rise to a gravitational well,
whose depth is proportional to the total number of gravitons present.
Since no other (matter) constituents appear in the model, we could view
it as a description of ``purely gravitational'' black holes, or the
approximation of the final state of gravitation collapse in which the initial matter
contribution has become subdominant with respect to the gravitons
themselves (in agreement with the huge gravitational entropy predicted
by Bekenstein for astrophysical black holes~\cite{bek}).
One point which remains unclear in Refs.~\cite{DvaliGomez} is whether
these systems in fact display a horizon, or trapping surface, as one
would expect in a ``standard'' black hole space-time.
\par
Before we delve into this point, let us briefly review the main argument in
Ref.~\cite{DvaliGomez}.
Note that we shall mostly use units with $c=1$, the Newton constant $\gn=\lp/\mpl$,
where $\lp$ and $\mpl$ are the Planck length and mass, respectively, and
$\hbar=\lp\,\mpl$.
These units make it apparent that $\gn$ converts mass into length,
thus providing a natural link between energy and size we shall assume
also holds at the quantum level~\cite{Cfuzzy}.
In the Newtonian approximation,
%, Membrado:1989ke,
%Balakrishna:1999sv,Nieuwenhuizen:2008zz,Nieuwenhuizen:2009px,
%Chavanis:2011cz}).},
we can assume a system of $N$ gravitons has total energy $M=N\,m$,
and that each graviton interacts with the others via the potential
\be
V_{\rm N}
\simeq
-\frac{\gn\,M}{r}
=
-\frac{\lp\,N\,m}{r\,\mpl}
\ ,
\label{Vn}
\ee
where the effective graviton mass $m$ is related to the characteristic quantum
mechanical size via the Compton/de~Broglie wavelength,
\be
\lambda_m
\simeq
\frac{\hbar}{m}
=
\lp\,\frac{\mpl}{m}
\ .
\ee
In fact, since gravitons can superpose, one expects them to give rise
to a ball of characteristic radius $r\simeq\lambda_m$ for sufficiently
large $N$.
For the following argument, it is then sufficient to assume the potential~\eqref{Vn}
becomes negligible for $r\gtrsim \lambda_m$ (for improved approximations, see
Refs.~\cite{Casadio:2013hja,mueckPT}), so that the potential energy of
each graviton interacting with the remaining $N-1$ gravitons is given by
\be
U_m(r)
&\!\!\simeq\!\!&
m\,V_{\rm N}(\lambda_m)
\nonumber
\\
&\!\!=\!\!&
-N\,\frac{\alpha\,\hbar}{\lambda_m}\,\Theta(\lambda_m-r)
\ ,
\label{Udvali}
\ee
where
\be
\alpha
=
\frac{\lp^2}{\lambda_m^2}
=
\frac{m^2}{\mpl^2}
\ee
is the effective gravitational coupling constant
and  $\Theta$ the Heaviside step function.
One can now see that there exist values of $N$ such that the system is
a black hole.
This happens when each graviton has just not enough kinetic energy
$E_{\rm K}\simeq m$ to escape the potential well, which yields
the marginally bound condition
\be
E_{\rm K} + U_m
\simeq
0
\ ,
\label{energy0}
\ee
equivalent to the ``maximal packing''  
\be
N\,\alpha = 1
\ .
\label{maxP}
\ee
From that point on, the ``blob'' of gravitons becomes a self-confined
object, whose effective boson mass and total mass scale
according to~\footnote{There is a large amount of works on self-gravitating bosons
in general relativity, which address the problem of gravitational stability
and black hole formation, and for which no analytical solution is available,
but where similar scaling relations can be found
(see, e.g.~Refs.~\cite{Ruffini:1969qy, Colpi:1986ye,Chavanis:2011cz}).}
\be
m
&\!\!\simeq\!\!&
\strut\displaystyle\frac{\mpl}{\sqrt{N}}
\label{mMax}
\\
M
&\!\!\simeq\!\!&
N\,m
\simeq
\sqrt{N}\,\mpl
\ .
\label{MMax}
\ee
Boson excitations can further lead to quantum depletion of the condensate
out of the ground state.
Such ``leaking'' of gravitons can be interpreted (at least in a first order
approximation) as the emission of Hawking radiation.
This kind of toy model is very intuitive and also gives an elegant
quantum mechanical description of black holes in term of the graviton
number $N$.
However, as we already mentioned, it leaves open the question 
whether the causal structure of space-time indeed contains a trapping
surface.
\par
We already noted that, in this model of ``purely gravitational'' black holes,
only gravitons are considered and there is no trace of (nor, apparently,
need of considering) the matter that initially collapsed and formed the black hole.
However, it is clear that, unless the black hole originated from a primordial
quantum fluctuation of the vacuum in the very early stages of the universe,
the only known mechanism that could possibly lead to such a final state
is the gravitational collapse of a star or other astrophysical source.  
The question then arises naturally as whether neglecting the role
of regular matter in the final state is a reliable approximation.
One could rather argue that matter always matters, and the
final state of gravitational collapse is not a ``pure'' black hole
like the one in Refs.~\cite{DvaliGomez}, if it is a black hole (in the
strict general relativistic sense) at all. 
This is the question one would eventually like to answer, although
it appears we are still quite far from that.
\par
In this work, we shall first review the picture and scaling relations~\eqref{mMax}
and \eqref{MMax} as proposed in Refs.~\cite{DvaliGomez} starting from the
relativistic field equation for scalar gravitons, but without assuming any specific
form for the necessary binding potential.
We shall start from the classical solution $\phi_{\rm c}$ of the Klein-Gordon
equation for a massless scalar field coupled to a static and
spherically symmetric matter source $J$.
Such a classical solution is reproduced, in the quantum theory,
by a coherent state obtained by superposing modes belonging
to a whole range of momenta $k>0$. 
However, if one assumes the source $J$ is determined by the scalar field
itself inside a finite spatial volume~\footnote{Let us remark that
this self-sourcing condition appears as a crucial aspect in ``classicalization''
of gravity, see Refs.~\cite{kuhnel2,dvaliCL}.
Of course, the existence of bound states localised inside a finite volume
requires an attractive self-interaction, of the kind one expects for
gravity~\cite{DvaliGomez}, and an example of which was studied
in Ref.~\cite{flassig}.},
namely if $J\sim \phi_{\rm c}$, one finds that (roughly) only one mode
$k^{-1}_{\rm c}\sim M$ is allowed.
According to the corpuscular model of Ref.~\cite{DvaliGomez}, this means the
quantum coherent state, representing the Newtonian
potential of a star, must have collapsed into a macroscopic quantum object
made of a large number $N$ of bosons in the same mode $k_{\rm c}$.
At this point, we shall finally be able to tackle the main issue of investigating
the presence of trapping surfaces by means of the formalism of the horizon
wave-function~\cite{Cfuzzy,GUPfuzzy,qhoop} for such quantum states in
the mode $k_{\rm c}$.
Our conclusion will be that there is indeed a horizon and that it is
of the expected classical size, for large $N$. 
We shall also consider a few generalisations of this state,
with diverse forms of ``quantum hair'', that could possibly be used
to model the Hawking radiation or the approach of collapsing matter
towards the horizon. 
We shall then see the uncertainty in the horizon radius always turns out
to be naturally related with the existence of leaking, or Hawking,
modes~\cite{pw}, and is thus determined by the Hawking temperature.
For larger black holes, this uncertainty in the horizon size clearly decreases,
in agreement with semiclassical expectations. 
It is then important to highlight that a similar result does not hold if one
tries to describe a black hole as a single very massive particle (that is, 
a system with $N=1$ and $m\gg \mpl$),
since in the latter case this uncertainty remains of the order of the horizon
size itself, regardless of the value of the black hole mass~\cite{GUPfuzzy}.
\par
In Section~\ref{starBH}, we first review the classical solutions of the
Klein-Gordon equation for a static source (thus equivalent to the
Poisson equation for Newtonian gravity), and the coherent states
that reproduce the classical Newtonian potential. 
We next sketch how the self-sustained state can generically arise in
this context.
The causal structure of the latter configuration is then analysed in
Section~\ref{BH} by means of the horizon wave-function of the system,
obtained from the spectral decomposition of a few different quantum
states of $N$ bosons distributed around the ground state $k_{\rm c}$.
One case in particular is discussed in which the distribution is
thermal, as is expected for the Hawking radiation.
We conclude with several comments and speculations in
Section~\ref{conc}.
\section{Massless scalar field toy model}
\label{starBH}
\setcounter{equation}{0}
We start by reviewing well-known general features of Newtonian gravity and
of the corpuscular model of black holes of Ref.~\cite{DvaliGomez} by means
of a toy scalar field.
This introductory material will allow us to highlight some of the main differences
between a (Newtonian) star and a black hole, and provide us with an approximate
quantum state for the subsequent analysis of the causal structure of space-time
in Section~\ref{BH}.
\par
Let us consider the Klein-Gordon equation for a real and massless scalar field
$\phi$ coupled to a scalar current $J$ in Minkowski space-time,
\be
\Box\phi(x)
=
q\,J(x)
\ ,
\label{eq:EOM}
\ee
where $\Box=\eta^{\mu\nu}\,\partial_\mu\,\partial_\nu$, the scalar field has
the standard canonical dimension of length$^{-1}$ and the coupling $q$ is
dimensionless for simplicity (appropriate dimensional factors will be introduced
in the final expressions).
We shall also assume the current is time-independent,
$\partial_0J=0$. 
In momentum space, with $k^\mu=(k^0,\mathbf{k})$, this implies that
\be
k^0\,\tilde{J}(k^\mu)
=
0
\ee
which is solved by the distribution
\be
\tilde{J}(k^\mu)
=
2\,\pi\,\delta(k^0)\,\tilde{J}(\mathbf{k})
\ ,
\ee
where $\tilde{J}^*(\mathbf{k})=\tilde{J}(-\mathbf{k})$.
For the spatial part, we further assume exact spherical symmetry,
so that our analysis is restricted to functions $f(\mathbf{x})=f(r)$,
with $r=|\mathbf{x}|$.
We can then introduce functions in momentum space according to
\be
\tilde{f}(k)
=
4\,\pi\int_0^{+\infty}{dr\,r^2\,j_0(kr)\,f(r)}
\ ,
\ee
where  
\be
j_0(kr)
=
\frac{\sin(kr)}{k\,r}
\ ,
\label{j0}
\ee
is a spherical Bessel function of the first kind and $k=|\mathbf{k}|$.
\subsection{Classical solutions}
\label{fic}
Classical spherically symmetric solutions of Eq.~\eqref{eq:EOM} can
be formally written as
\be
\phi_{\rm c}(r)
=
q\,\Box^{-1}
J(r)
\ ,
\label{eq:EOMmom}
\ee
and are given in momentum space by
\be
\tilde{\phi}_{\rm c}(k)
=
-q\,\frac{\tilde{J}(k)}{k^2}
\ .
\label{eqkc}
\ee
For example, if the current has Gaussian support,
\be
J(r)
=
\frac{e^{-r^2/(2\,\sigma^2)}}{(2\,\pi\,\sigma^2)^{3/2}}
\ ,
\label{eq:Gauss}
\ee
so that
\be
\tilde{J}(k)
=
e^{-k^2\sigma^2/2}
\ ,
\ee
the corresponding classical scalar solution is given by
\be
\phi_{\rm c}(r)
&\!\!=\!\!&
-\frac{q}{2\,\pi^2}\int_0^{+\infty}{\d k\,
j_0(kr)\,e^{-k^2\sigma^2/2}}
\nonumber
\\
&\!\!=\!\!&
-\frac{q}{4\,\pi\,r}\,
\mathrm{erf}{\left(\frac{r}{\sqrt{2}\,\sigma}\right)}
\ .
\ee
The above expression outside the source $J$, or for $r\gg\sigma$,
reproduces the classical Newtonian potential~\eqref{Vn}, namely 
\be
V_{\rm N}
=
\frac{4\,\pi}{q}\,\gn\,M\,\phi_{\rm c}
\simeq
-\frac{\gn\,M}{r}
\ ,
\label{V_N}
\ee
where we introduced the suitably dimensioned factor.
\subsection{Quantum coherent states}
\label{Qcoe}
In the quantum theory, the classical configurations~\eqref{eq:EOMmom}
are replaced by coherent states.
This can be easily seen from the normal-ordered quantum Hamiltonian density
in momentum space,
\be
\hat{\mathcal{H}}
=
k\,\hat a'^\dagger_k\,\hat a'_k+\tilde{\mathcal{H}}_g
\ ,
\label{eq:hamquant}
\ee
where $\tilde{\mathcal{H}}_g$ is the ground state energy density, 
\be
\tilde{\mathcal{H}}_g
=
-q^2\,\frac{|\tilde{J}(k)|^2}{2\,k^2}
\ ,
\ee
and we shifted the standard ladder operators according to
\be
\hat a'_k
=
\hat a_k+q\,\frac{\tilde{J}(k)}{\sqrt{2\,k^3}}
\ .
\label{eq:laddshift}
\ee
The source-dependent ground state $\Ket{g}$ is annihilated by the
shifted annihilation operator,
\be
\hat a'_k
\Ket{g}
=
0
\ ,
\ee
and is a coherent state in terms of the standard field vacuum,
\be
\hat a_k
\Ket{g}
=
-q\,\frac{\tilde{J}(k)}{\sqrt{2\,k^3}}
\Ket{g}
=
g(k)
\Ket{g}
\ ,
\ee
where $g=g(k)$ is thus an eigenvalue of the shifted annihilation operator.
This implies
\be
\Ket{g}
=
e^{-N/2}\,
{\rm {exp}}\left\{\int{\frac{k^2\,\d k}{2\,\pi^2}\,g(k)\,\hat a^\dagger_k}\right\}
\Ket{0}
\ ,
\label{ketg}
\ee
where $N$ denotes the expectation value of the number of quanta
in the coherent state,
\be
N
&\!\!=\!\!&
\int{\frac{k^2\,\d k}{2\,\pi^2}\,
\Bra{g}\hat a^\dagger_k\,\hat a_k\Ket{g}}
\nonumber
\\
&\!\!=\!\!&
\int{\frac{k^2\,\d k}{2\,\pi^2}\,|g(k)|^2}
\nonumber
\\
&\!\!=\!\!&
\frac{q^2}{(2\,\pi)^2}
\int \frac{\d k}{k}\,
|\tilde{J}(k)|^2
\ ,
\label{expN}
\ee
from which we can read off the occupation number
\be
n_k
=
\left(\frac{q}{2\,\pi}\right)^2
\frac{|\tilde{J}(k)|^2}{k}
\ .
\ee
It is straightforward to verify that the expectation value of the field in the
state $\Ket{g}$ coincides with its classical value,
\be
\Bra{g}\hat {\phi}_k\Ket{g}
&\!\!=\!\!&
\frac{1}{\sqrt{2\,k}}
\Bra{g}
\left(\hat a_k+\hat a^\dagger_{-k}\right)
\Ket{g}
\nonumber
\\
&\!\!=\!\!&
\frac{1}{\sqrt{2\,k}}
\Bra{g}
\left(\hat a'_k+\hat a'^\dagger_{-k}\right)
\Ket{g}
-q\,\frac{\tilde{J}(k)}{k^2}
\nonumber
\\
&\!\!=\!\!&
\tilde \phi_{\rm c}(k)
\ ,
\label{expPc}
\ee
thus $\ket{g}$ is a realisation of the Ehrenfest theorem.
\par
It is interesting to recall that the state $\Ket{g}$ and the number $N$
are not mathematically well-defined in general:
Eq.~\eqref{expN} is UV divergent if the source
has infinitely thin support, and IR divergent if the source
contains modes of vanishing momenta (which would only be physically
consistent with an eternal source).
The UV issue can be cured, for example, by using a Gaussian distribution
like the one in Eq.~\eqref{eq:Gauss},
while the IR divergence can be naturally eliminated if the scalar field is 
massive or the system is enclosed within a finite volume
(so that allowed modes are also quantised).
These details are however of little importance for what follows and
we shall therefore not indulge in them here.
\subsection{Stars and self-sustained scalar states}
\label{BHstar}
The system we have considered so far could represent a ``star'',
that is a classical lump of ordinary matter with density
\be
\rho=M\,J
\ ,
\label{rhoM}
\ee
where $M$ is the total (proper) energy of the star.
The Newtonian potential energy for this star is of course given by $U_{M}=M\,V_{\rm N}$,
so that $\phi_{\rm c}$ is accordingly determined by Eq.~\eqref{V_N} and the quantum state
of $\hat \phi$ by Eq.~\eqref{expPc}.
Note that, for very large mass $M$, the scalar gravitons in the coherent state $\ket{g}$
are mostly found with an energy around $m\sim U_{M}/N$ and their total number is
$N\sim M^2$, in agreement with Eq.~\eqref{MMax}~\cite{DvaliGomez}.
It is interesting to note that this scaling relation for the total mass of the self-gravitating
system holds both for a star and in the black hole regime, whereas Eq.~\eqref{mMax}
for the graviton's energy holds {\em only\/} in the black hole regime (since the total
Newtonian potential energy $U_{M}\simeq N\,m\ll M$ for a regular star).
\par
Let us instead assume there exists a regime in which the matter contribution is negligible,
and the source $J$ in the r.h.s.~of Eq.~\eqref{eq:EOM} is thus provided by the gravitons
themselves~\cite{DvaliGomez}, (at least) inside a finite spatial volume $\mathcal{V}$.
As we mentioned in the Introduction, this confinement is a crucial feature for
``classicalization'' of gravity~\cite{dvaliCL}, and necessarily requires an attractive
self-interaction for the scalar field to admit bound states.
Of course, one expects this to happen for full-fledged gravity and, in the semiclassical
regime, to lead to a curved space-time metric (inside the volume $\mathcal{V}$).
The self-interaction, at least in this regime, could then be effectively described by
a modified D'Alembertian in Eq.~\eqref{eq:EOM}.
Since the details of the (otherwise necessary) confining mechanism are not
relevant for our analysis, we can just assume the momentum modes in Eq.~\eqref{j0}
are replaced by those in the appropriate curved space-time and that Eq.~\eqref{eqkc}
in momentum space still holds. 
In other words, we just require that the energy density~\eqref{rhoM} sourcing the
evolution of each scalar, is equal to minus the average total potential energy
$N\,U_m/\mathcal{V}$~\footnote{The total potential energy of each graviton is
proportional to $(N-1)\,U_m$, but since we are interested in the large $N$
case, we approximate $N-1\simeq N$.}, where $\mathcal{V}=4\,\pi\,R^3/3$
is the spatial volume where this source has support.
This is just the same as the marginally bound condition~\eqref{energy0} used
in Refs.~\cite{DvaliGomez} with $N\,E_{\rm K}\sim J$, and implies that
\be
J
\simeq
-\frac{3\,N\,\gn\,m}{q\,R^3}\,\phi_{\rm c}
\ ,
\label{Jg}
\ee
inside the volume $\mathcal{V}$, where $m$ is again the energy of each scalar
graviton and $N$ their total number.
Upon replacing this condition into Eq.~\eqref{eqkc},
we straightforwardly obtain
\be
\frac{3\,N\,\gn\,m}{R^3\,k^2}
=
\frac{3\,\Rh}{2\,R^3\,k^2}
\simeq
1
\ ,
\ee
where
\be
\Rh=2\,\gn\,M
\label{Rh}
\ee
is the classical Schwarzschild radius associated with the
total mass $M=N\,m$.
This suggest that, unlike the Newtonian potential generated by an ordinary
matter source, a self-sustained system should contain only the
modes with momentum numbers $k=k_{\rm c}$ such that
\be
R\,k_{\rm c}
\simeq 
\sqrt{\frac{\Rh}{R}}
\ ,
\label{kc}
\ee
where we dropped a numerical coefficient of order one given the qualitative
nature of our analysis.
Ideally, this means that, if the scalar gravitons represent the main gravitating source,
the quantum state of the system must be given in terms of just one mode
$\phi_{k_{\rm c}}$.
A coherent state of the form in Eq.~\eqref{ketg} cannot thus be built,
and the relation~\eqref{expN} between $N$ and the source momenta
does not apply here.
Instead, for $N\gg 1$, all scalars will be in the same state $\ket{k_{\rm c}}$.
\par
For an ordinary star, the typical size $R\gg \Rh$ and $k_{\rm c}\ll R^{-1}$.
The corresponding de~Broglie length $\lambda_{\rm c}\simeq k^{-1}_{\rm c}\gg R$, 
which would conflict with our assumption that the field be identified as
a gravitating source  only within a region of size $R$.
However, if we consider the ``black hole limit'' in which $R\sim\Rh$,
and recalling that $m=\hbar\,k$, we immediately obtain, again dropping 
a numerical coefficient of order one, 
\be
1
\simeq
\gn\,M\,k_{\rm c}
=
N\,\frac{m^2}{\mpl^2}
\ ,
\label{1=N}
\ee
which leads to the two scaling relations~\eqref{mMax} and \eqref{MMax}, namely
$m=\hbar\,k_{\rm c}\simeq \mpl/\sqrt{N}$ 
and a consistent de~Broglie length $\lambda_m\simeq\lambda_{\rm c}\simeq \Rh$.
An ideal system of self-sustained scalars should then be in the quantum state
$\ket{k_{\rm c}}$~\footnote{For more details about the identification of this state
as a BEC, see for example, Ref.~\cite{flassig}.}
 with a spatial size that hints at the system as a black
hole~\cite{DvaliGomez}.
In order to substantiate this last part of the sentence, one should however show
that there is a horizon, or at least a trapping surface, in the given space-time.
\par
The standard procedure to show the existence of trapping surfaces requires
knowing explicitly the metric that solves the semiclassical Einstein field equations
with the prescribed source, the latter being represented by the expectation value
of the appropriate stress-energy tensor.
One therefore also needs the explicit form of the momentum modes we implicitly
assumed lead to Eq.~\eqref{eqkc}.
However, the problem of self-gravitating scalar fields in general relativity
has been know for decades~\cite{Ruffini:1969qy} and analytical solutions
have yet to be found~\cite{Colpi:1986ye,Chavanis:2011cz}.
Moreover, it is not {\em a priori\/} guaranteed that such a semiclassical approach
remains valid for the type of quantum sources we are dealing with,
since the quantum fluctuations of the source could be large enough to spoil
the possibility of employing a curved background geometry to describe the region where
the source is located~\footnote{Let us remark that, without a curved background geometry,
the very definition of a trapping surface becomes conceptually challenging~\cite{Cfuzzy}.}.
In fact, one expects this standard approach in general holds at large distance
from the source, and yields the proper (post-)Newtonian approximation~\cite{DvaliGomez},
in a region far from where trapping surfaces are likely to be located.
This is precisely the reason a formalism for describing the gravitational
radius of any quantum system was introduced in Refs.~\cite{Cfuzzy,GUPfuzzy,qhoop},
as we shall review shortly.
\par
Before we do so, let us remark that the above argument leading to Eq.~\eqref{1=N}
does not need the scalar field $\phi$ to vanish (or be negligible) outside the region
of radius $\Rh$.
This would in fact imply there is no outer Newtonian potential,
which is instead quite contrasting with the idea of a gravitational
source.
All we need in order to recover the classical outer Newtonian potential
$V_{\rm N}\sim\phi$ is to relax the condition~\eqref{Jg}
for $r\gtrsim \Rh$ (where $J\simeq 0$), and properly match the
(expectation) value of $\hat\phi$ with the Newtonian $\phi_{\rm c}$ from
Eq.~\eqref{V_N} at $r\gtrsim \Rh$.
In this matching procedure, at least for $r\gg\Rh$ and $N\gg 1$, one then expects to
recover the classical description for which the only relevant information is
the total mass $M$ of the black hole. 
This can already be seen from the classical analysis of the outer ($r\gg\sigma\sim\Rh$)
Newtonian scalar potential in Section~\ref{fic}, and its quantum counter-part
in Section~\ref{Qcoe}, but also in the alternative description of gravitational scattering.
It has in fact been known for a long time that the geodesic motion in the
post-Newtonian expansion of the Schwarzschild metric can be reproduced
by tree-level Feynman diagrams with graviton exchanges between a test probe
and a (classical) large source~\cite{duff}~\footnote{For a similarly non-geometric
derivation of the action of Einstein gravity, see Ref.~\cite{deser}.}.
In this calculation, again for $r\gg\Rh$, the source is just described by its
total mass $M$, and quantum effects should then be suppressed by factors of
$1/N$~\cite{DvaliGomez}. 
\section{Horizon of self-sustained states}
\label{BH}
\setcounter{equation}{0}
%
%
%
%
%
%\subsection{Horizon}
%
In the above review, there is no explicit evidence of a non-trivial causal structure.
As we already mentioned, in order to show that a system of $N\gg 1$ scalar gravitons
is a black hole in the usual sense, we must be able to identify a radius $r\sim \Rh$
as the actual horizon.
In order to do so, we employ the horizon wave-function introduced in
Ref.~\cite{Cfuzzy} (see Refs.~\cite{GUPfuzzy,qhoop} for more details).
\par
This formalism can be applied to the quantum mechanical state $\psi_{\rm S}$
of any system {\em localised in space\/} and {\em at rest\/} in the chosen
reference frame.
Having defined suitable Hamiltonian eigenmodes,
$
\hat H\,\ket{\psi_E}=E\,\ket{\psi_E}
$,
where $H$ can be specified depending on the model we wish to consider,
the state $\psi_{\rm S}$ can be decomposed as
\be
\ket{\psi_{\rm S}}
=
\sum_E\,C(E)\,\ket{\psi_E}
\ .
\label{CE}
\ee
If we further assume the system is {\em spherically symmetric\/},
we can invert the expression of the Schwarzschild radius,
\be
\rh
=
2\,\gn\,E
\ ,
\label{rh}
\ee
in order to obtain $E$ as a function of $\rh$.
We then define the {\em horizon wave-function\/} as
\be
\psi_{\rm H}(\rh)
\propto
C\left(\mpl\,{\rh}/{2\,\lp}\right)
\ ,
\ee
whose normalisation is finally fixed in the inner product
\be
\pro{\psi_{\rm H}}{\phi_{\rm H}}
=
4\,\pi\,\int_0^\infty
\psi_{\rm H}^*(\rh)\,\phi_{\rm H}(\rh)\,\rh^2\,\d \rh
\ .
\ee
We interpret $\psi_{\rm H}$ simply as the wave-function yielding
the probability $P_{\rm H}(\rh)=4\,\pi\,\rh^2\,|\psi_{\rm H}(\rh)|^2$
that we would detect a gravitational radius $r=\rh$
associated with the given quantum state $\psi_{\rm S}$.
Such a radius generalises the classical concept of the Schwarzschild
radius of a spherically symmetric distribution of matter and is necessarily
``fuzzy'', like the position and energy of the particle itself~\cite{Cfuzzy,GUPfuzzy}.
The probability density that the system lies inside its own gravitational
radius $r=\rh$ will next be given by the conditional expression
\be
P_<(r<\rh)
=
P_{\rm S}(r<\rh)\,P_{\rm H}(\rh)
\ ,
\label{PrlessH}
\ee
where
$
P_{\rm S}(r<\rh)
=
4\,\pi\,\int_0^{\rh}
|\psi_{\rm S}(r)|^2\,r^2\,\d r
$
is the probability that the system is inside a sphere of radius $r=\rh$.
Finally, the probability that the system described by the wave-function $\psi_{\rm S}$
is a black hole will be obtained by integrating~\eqref{PrlessH} over all possible
values of the radius,
\be
P_{\rm BH}
=
\int_0^\infty P_<(r<\rh)\,\d \rh
\ .
\label{PBH}
\ee
When $P_{\rm BH}\simeq 1$, the system is very likely found inside its own
gravitational radius, which therefore turns into a trapping surface (or, loosely speaking,
a horizon), and can be (at least temporarily) viewed as a black hole.  
The above general formulation can be easily applied to a particle described by a
spherically symmetric Gaussian wave-function, for which one obtains a vanishing 
probability that the particle is a black hole when its mass is much smaller than $\mpl$
(and uncertainty in position $\lambda_m\gg\lp$)~\cite{Cfuzzy,GUPfuzzy}.
However, the uncertainty in the horizon size for a single particle with trans-Planckian
mass $M\gg\mpl$ turns out to be
\be
\Delta\rh
\sim
\expec{\hat r_{\rm H}}
\sim
\Rh
\ ,
\label{Mmp}
\ee
which clearly shows that this system cannot represent a large, semiclassical
black hole.
\par
The main difference with respect to a single very massive particle~\cite{Cfuzzy,GUPfuzzy}
is that our system is now composed of a very large number $N$ of particles of very small
effective mass $m\ll \mpl$ (thus very large de~Broglie length, $\lambda_m\gg\lp$).
According to Refs.~\cite{GUPfuzzy}, they cannot individually form (light) black holes,
however the generalisation of the formalism to a system of $N$ such components will
enable us to show that the total energy $E=M$ is indeed sufficient to create a
proper horizon.
\subsection{Hairless black hole}
\label{idealBH}
\begin{figure}[t]
\centering
\includegraphics[width=14cm]{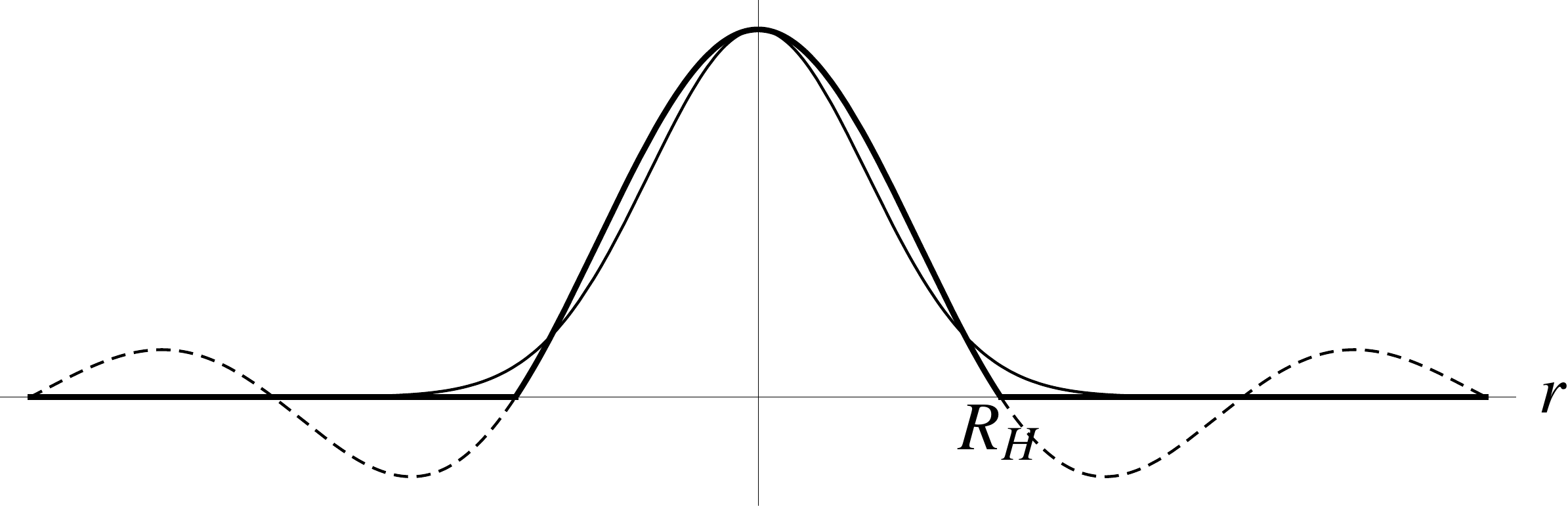}
\caption{Scalar field mode of momentum number $k_{\rm c}$:
ideal approximation in Eq.~\eqref{psiS0} (thick solid line) compared
to exact $j_0(k_{\rm c} r)$ (dashed line).
The thin solid line represents a Gaussian distribution of the kind considered
in Ref.~\cite{Casadio:2013hja}.
\label{ideal}}
\end{figure}
Let us first consider the highly idealised case in which Eq.~\eqref{kc} admits
precisely one mode, defined by 
\be
k_{\rm c}
=
\frac{\pi}{\Rh}
=
\frac{\pi}{2\,\sqrt{N}\,\lp}
\ ,
\ee
so that $\phi_{k_{\rm c}}(\Rh)\simeq j_0(k_{\rm c}\,\Rh)=0$ and the scalar field vanishes
outside of $r=\Rh$,
\be
\psi_{\rm S}(r_i)
=
\pro{r_i}{k_{\rm c}}
=
\left\{
\begin{array}{ll}
\mathcal{N}_{\rm c}\,j_0(k_{\rm c} r_i)
&
{\rm for}\ 
r<\Rh
\\
\\
0
&
{\rm for}\ 
r>\Rh
\ ,
\end{array}
\right.
\label{psiS0}
\ee
where $\mathcal{N}_{\rm c}=\sqrt{\pi/2\,\Rh^3}$ is a normalisation factor such that
\be
4\,\pi\,\mathcal{N}_{\rm c}^2
\int_0^{\Rh}
|j_0(k_{\rm c} r)|^2
\,r^2\,\d r
=
1
\ .
\ee
The above approximate mode is plotted in Fig.~\ref{ideal}, where it is also
compared with a Gaussian distribution of the kind considered
in Ref.~\cite{Casadio:2013hja}, which appears qualitatively very similar. 
\par
We have already commented in the previous Section that a scalar field
which vanishes everywhere outside $r=\Rh$ is actually inconsistent with the presence
of an outer Newtonian potential, but let us put this fact aside momentarily.
The wave-function of the system of $N$ such modes is 
the (totally symmetrised) product of $N\sim M^2$ equal modes [see the order
$\gamma^0$ term in Eq.~\eqref{NGausG}, with $\ket{m}=\ket{k_{\rm c}}$],
\be
\psi_{\rm S}(r_1,\ldots,r_N)
=
\frac{\mathcal{N}_{\rm c}^N}{N!}\,
\sum_{\{\sigma_i\}}^N\,
\prod_{i=1}^N
\,
j_0(k_{\rm c} r_i) 
\ ,
\ee
where the sum is over all the permutations $\{\sigma_i\}$ of the $N$ excitations.
This is obviously an energy eigenstate,
\be
\hat H\,\psi_{\rm S}
=
N\,\hbar\,k_{\rm c}\,\psi_{\rm S}
=
M\,\psi_{\rm S}
\ ,
\ee
where $\hat H=\sum_i\,\hat H_i=\sum_i\,\hbar\,\hat k_i$ is the total Hamiltonian for
$N$ free massless scalars in momentum space.
The only non-vanishing coefficient in the spectral decomposition is then
given by $C(E)=1$, for $E=N\,\hbar\,k_{\rm c}=M$, corresponding to a probability
density for finding the horizon size between $\rh$ and $\rh+\d\rh$  
\be
\d P_{\rm H}(\rh)
&\!\!=\!\!&
4\,\pi\,
\rh^2\,
\left|\psi_{\rm H}(\rh)\right|^2\,
\d\rh
\nonumber
\\
&\!\!=\!\!&
\delta(\rh-\Rh)\,
\d\rh
\ .
\ee
This result, along with the fact that all $N$ excitations in the mode $k_{\rm c}$
are confined within the radius $\Rh\simeq\lambda_{\rm c}$,
\be
P_{\rm S}(r_i<\Rh)
=
4\,\pi\,\mathcal{N}_{\rm c}^2
\int_0^{\Rh}
|j_0(k_{\rm c}\,r)|^2
\,r^2\,\d r
=
1
\ ,
\ee
immediately leads to the conclusion that the system is indeed a black hole,
\be
P_{\rm BH}
&\!\!\simeq\!\!&
4\,\pi\,\mathcal{N}_{\rm c}^2
\int_0^\infty
\d\rh\,
\delta(\rh-\Rh)
\int_0^{\rh}
|j_0(k_{\rm c}\,r)|^2
\,r^2\,\d r
\nonumber
\\
&\!\!=\!\!&
P_{\rm S}(r<\Rh)
=
1
\ .
\ee
\par
In the above ``ideal'' approximation~\eqref{psiS0}, the horizon would exactly be located
at its classical radius,
\be
\expec{\hat r_{\rm H}}
\equiv
\bra{\psi_{\rm H}} \hat r_{\rm H}\ket{\psi_{\rm H}}
=\Rh
\ ,
\label{expRh}
\ee
with absolutely negligible uncertainty, $\Delta\rh\simeq 0$, where
\be
\Delta\rh^2
\equiv
\bra{\psi_{\rm H}}\left(\hat r_{\rm H}^2-\Rh^2\right)\ket{\psi_{\rm H}}
\ .
\ee
A zero uncertainty in $\expec{\hat r_{\rm H}}$ is not a sound
result, which likely parallels the description of a macroscopic black hole
as a pure quantum mechanical state built out of the single-particle 
wave-functions in Eq.~\eqref{psiS0}.
Moreover, we recall again that a non-vanishing scalar field at $r>\Rh$
is also necessary in order to reproduce the expected outer Newtonian potential.
\subsection{Black hole with quantum hair}
\label{BHhair}
It is reasonable that a more realistic macroscopic black hole of the kind 
we consider here (with $N\gg 1$) is not just an energy eigenstate with $k=k_{\rm c}$
but contains more modes.
In particular, we expect the mode with $k= k_{\rm c}$ forms a discrete spectrum
(which is  tantamount to assuming $k_{\rm c}$ is the minimum allowed momentum,
in agreement with the idea of a BEC of gravitons), and must be treated separately.
Modes with $k> k_{\rm c}$ would instead be able to ``leak out'' (roughly representing
the Hawking flux, as we shall see), and form a continuous spectrum, which will
turn out to be responsible for the fuzziness in the horizon's location.
\par
\begin{figure}[t]
\centering
\includegraphics[width=14cm,height=7cm]{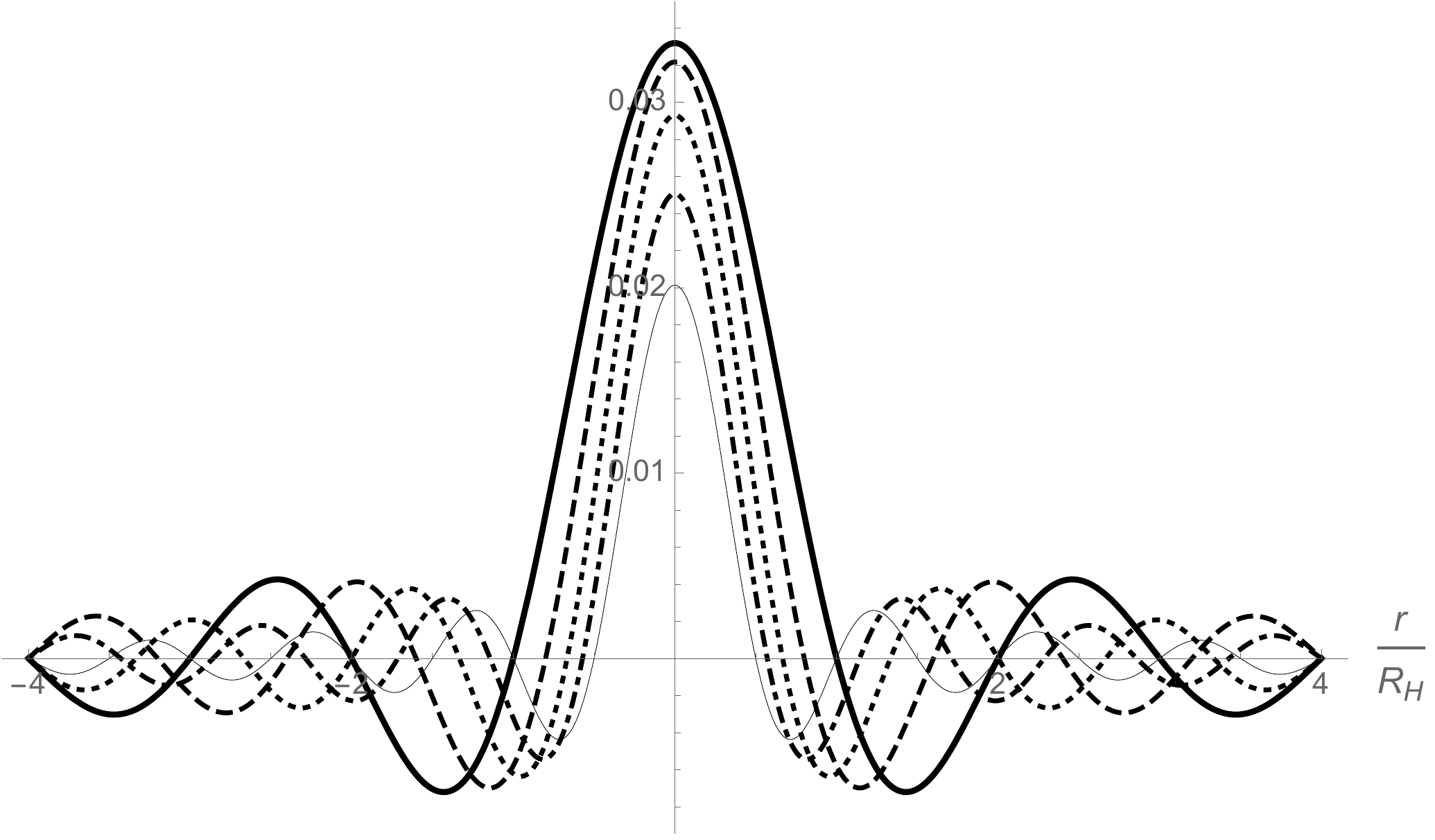}
\caption{Modes of momentum number $k=k_{\rm c}$ (thick solid line),
$k=(5/4)\,k_{\rm c}$ (dashed line),
$k=(3/2)\,k_{\rm c}$ (dotted line),
$k=(7/4)\,k_{\rm c}$ (dash-dotted line)
and
$k=2\,k_{\rm c}$ (thin solid line).
The relative weight is determined according to Eq.~\eqref{psi_i}.
\label{modes}}
\end{figure}
For the sake of employing a calculable function, let us assume here
the continuous distribution in momentum space of each of the $N$
scalar states is given by half a Gaussian peaked around $k_{\rm c}$
(see Fig.~\ref{modes} for a few modes above $k_{\rm c}$),
\be
\ket{\psi_{\rm S}^{(i)}}
=
\mathcal{N}_\gamma
\left(
\ket{k_{\rm c}}
+
\gamma
\int_{k_{\rm c}}^\infty
\frac{\sqrt{2}\,\d k_i}{\sqrt{\Delta_i\,\sqrt{\pi}}}\,
e^{-\frac{\hbar^2(k_i-k_{\rm c})^2}{2\,\Delta_i^2}}
\ket{k_i}
\right)
\ ,
\label{psi_i}
\ee
where $i=1,\ldots,N$, the ket $\ket{k}$ denotes the eigenmode of eigenvalue $k$,
and
\be
\mathcal{N}_\gamma
=
\left(1+\gamma^2\right)^{-1/2}
\ee
is a global normalisation factor.
The parameter $\gamma$ is a real and dimensionless coefficient which weighs
the relative probability of finding the particle in the continuous part of the spectrum
with respect to the same particle being in the discrete ground state $\phi_{k_{\rm c}}$,
and we shall see later on that it plays a major role in our analysis.
In particular, one should note that we are here assuming $\gamma$ does not depend
on $N$.
We shall also assume the width $\Delta_i=m\simeq M/N\simeq \mpl/\sqrt{N}$,
as follows from the typical mode spatial size $k_{\rm c}^{-1}\sim \sqrt{N}\,\lp$,
and is the same for all particles~\footnote{This assumption will help to simplify
the calculations, although it would be perhaps more realistic to assume a
different width for each mode $k_i$.}.
Since $m=\hbar\,k_{\rm c}$ and $E_i=\hbar\,k_i$, one can also write
\be
\ket{\psi_{\rm S}^{(i)}}
=
\mathcal{N}_\gamma
\left(
\ket{m}
+
\gamma
\int_{m}^\infty
\frac{\sqrt{2}\,\d E_i}{\sqrt{m\,\sqrt{\pi}}}\,
e^{-\frac{(E_i-m)^2}{2\,m^2}}
\ket{E_i}
\right)
\ .
\label{psiEi}
\ee
The total wave-function will be given by the symmetrised product of $N$ such
states [see Eq.~\eqref{NGaus}],
and one can then identify two regimes, depending on the value of $\gamma$
(see Appendix~\ref{mixS} for all the details).
\par
For $\gamma\ll 1$, to leading order in $\gamma$, one finds that the spectral
coefficient for $E\ge M$ is given by the contribution of the discrete quantum state
(corresponding to all of the $N$ particles in the mode $k_{\rm c}$) plus the
contribution with just one particle in the continuum [see Eq.~\eqref{CE=M} and
the first term in the r.h.s.~of Eq.~\eqref{CE3}],
\be
C(E\ge M)
\simeq
\mathcal{N}_\gamma
\left[
\delta_{E,M}
+
\gamma
\left(\frac{2}{m\,\sqrt{\pi}}\right)^{1/2}
e^{-\frac{(E-M)^2}{2\,m^2}}
\right]
\ ,
\ee
where $\delta_{A,B}$ is a Kronecker delta for the discrete part of the spectrum,
and the width $m\sim \mpl/\sqrt{N}$ is precisely the typical energy of Hawking quanta
emitted by a black hole of mass $M\simeq \sqrt{N}\,\mpl$.
It is then easy to compute the expectation value of the energy to next-to-leading
order for large $N$ and small $\gamma$,
\be
\expec{E}
&\!\!\simeq\!\!&
\mathcal{N}_\gamma^2
\left(
M+
\int_M^\infty
E\,C^2(E)\,\d E
\right)
\nonumber
\\
&\!\!\simeq\!\!&
\sqrt{N}\,\mpl
\left(1+\frac{\gamma^2/\sqrt{\pi}}{1+\gamma^2}\,\frac{1}{N}
\right)
\nonumber
\\
&\!\!\simeq\!\!&
\sqrt{N}\,\mpl
\left(1+\frac{\gamma^2}{\sqrt{\pi}\,N}
\right)
\ ,
\ee
and its uncertainty 
\be
\Delta E
=
\sqrt{\expec{E^2}-\expec{E}^2}
\simeq
\frac{\gamma\,\mpl}{\sqrt{2\,N}}
\ .
\ee
Putting the above two expressions together, we obtain the ratio
\be
\frac{\Delta E}{\expec{E}}
\simeq
\frac{\gamma}{\sqrt{2}\,N}
\ ,
\label{DEg}
\ee
where we just kept the leading order in the large $N$ expansion and neglected
terms of higher order in $\gamma$.
From the expression of the Schwarzschild radius~\eqref{rh}, or $\rh=2\,\lp\,E/\mpl$,
we then immediately obtain $\expec{\hat r_{\rm H}}\simeq \Rh$, with $\Rh$ given in
Eq.~\eqref{Rh}, and
\be
\frac{\Delta\rh}{\expec{\hat r_{\rm H}}}
\sim
\frac{1}{N}
\ ,
\ee
which vanishes rather fast for large $N$.
This case could hence describe a macroscopic BEC black hole with (very) little
quantum hair, in agreement with Refs.~\cite{DvaliGomez}, thus overcoming
the problem of the excessive large fluctuations~\eqref{Mmp} associated with
a single massive particle.
\par
Since this ``hair'' is indeed expected to be the Hawking radiation field,
we shall make the connection with thermal Hawking radiation
more explicit in Section~\ref{ThBH}, and obtain essentially the same estimate
of the horizon uncertainty therein.
\subsection{Quantum hair with no black hole}
\label{noBH}
For $\gamma\gtrsim 1$ and $N\gg 1$, one obtains the distribution in energy
is dominated by all of the $N$ particles in the continuum [see the last term in
the r.h.s.~of Eq.~\eqref{CE3}], so that the ground state $\phi_{k_{\rm c}}$ is
actually depleted (or was never occupied).
\par
Since the coefficient $\gamma^N$, as well as any other overall factors,
can be omitted in this case, one finds 
\be
C(E\ge M)
\simeq
\int_m^\infty
\d E_1\cdots
\int_m^\infty
\d E_N\,
\exp\left\{-\sum_{i=1}^N\frac{(E_i-m)^2}{2\,m^2}\right\}
\,\delta\left(E-\sum_{i=1}^N E_i\right)
\ ,
\label{CE1}
\ee
along with $C(E<M)\simeq 0$.
Note that the mass $M=N\,m$ is still to be viewed as the minimum energy of
the system corresponding to the ``ideal'' black hole with all of the
$N$ particles in the ground state $\ket{k_{\rm c}}$.
For $N=M/m\gg1$, this spectral function is estimated analytically in
Appendix~\ref{contCE}, and is given by~\footnote{See also Appendix~\ref{numCE},
where we compare Eq.~\eqref{CE>M} with a numerical estimate using a standard Monte Carlo
method.} 
\be
C(E\ge M)
\simeq
\sqrt{\frac{2}{\pi\,m^3}}
\,(E-M)\,
e^{-\frac{(E-M)^2}{4\,m^2}}
\ ,
\label{CE>M}
\ee
which is peaked slightly above $E\simeq M=N\,m$, with a width
$\sqrt{2}\,\Delta_i\sim m$, so that the (normalised)
expectation value
\be
\expec{E}
%&\!\!\simeq\!\!&
\simeq
\int_M^\infty
E\,C^2(E)\,\d E
%\nonumber
%\\
%&\!\!\simeq\!\!&
=
M+2\,\sqrt{\frac{2}{\pi}}\,m
\ ,
\ee
in agreement with the fact that we are now considering a system built out
of continuous modes whose energy must be (slightly) larger than $m$.
For $N\gg 1$, however, $\expec{E}=M\,[1+\mathcal{O}(N^{-1})]$, and
the energy quickly approaches the minimum value $M$, as is confirmed
by the uncertainty
\be
\Delta E
\simeq
\sqrt{\frac{3\,\pi-8}{\pi}}\,m
\ ,
\ee
or $\Delta E\sim N^{-1/2}$.
\par
\begin{figure}[t]
\centering
\includegraphics[width=14cm]{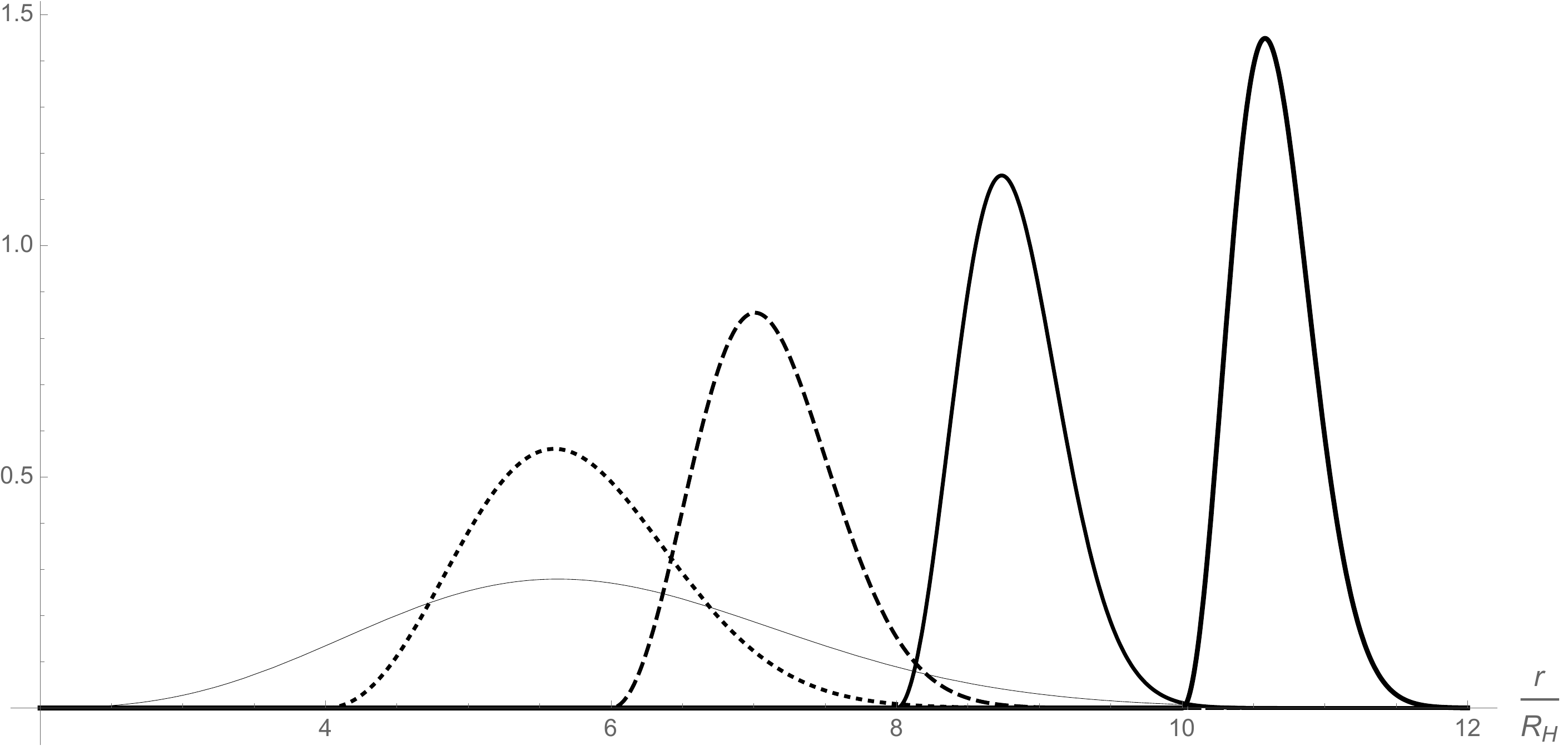}
\caption{Probability density of finding the horizon with radius $\rh$ for
$N=1$ ($\Rh=2\,\lp$; thin solid line),
$N=4$ ($\Rh=4\,\lp$; dotted line),
$N=9$ ($\Rh=6\,\lp$; dashed line),
$N=16$ ($\Rh=8\,\lp$; solid line)
and
$N=25$ ($\Rh=10\,\lp$; thick solid line).
The curves clearly become narrower the larger $N$. 
\label{plotRh}}
\end{figure}
The corresponding horizon wave-function is again obtained by simply
recalling that $\rh=2\,\lp\,E/\mpl$, and is approximately given by
\be
\psi_{\rm H}(\rh\ge 2\,\sqrt{N}\,\lp)
\simeq
%\mathcal{N}_N
\left(\rh-2\,\sqrt{N}\,\lp\right)
\,
e^{-\frac{\left(\rh-2\,\sqrt{N}\,\lp\right)^2}{16\,\lp^2/N}}
\ ,
\ee
%where $\mathcal{N}_N$ is a normalisation factor,
and $\psi_{\rm H}(\rh< 2\,\sqrt{N}\,\lp)\simeq 0$.
The probability density of finding the horizon with a radius between $\rh$ and $\rh+\d\rh$
is plotted in Fig.~\ref{plotRh} for a few values of $N$.
It is clear that for $N\sim 1$, the uncertainty in the horizon location would be large,
but it decreases very fast for increasing $N$.
Accordingly, the (unnormalised) expectation value
\be
\expec{\hat r_{\rm H}}
\simeq
2\,\sqrt{N}\,\lp
\left(
1 
+\sqrt{\frac{2}{\pi}}\,\frac{2}{N}
\right)
=
\Rh
\left[
1 
+
\mathcal{O}(N^{-1})
\right]
\ ,
\ee
which approaches the horizon radius of the ideal black hole,
$\Rh=2\,\sqrt{N}\,\lp$, for large $N$.
In agreement with previous comments about the distribution in energy,
the position of the horizon has an uncertainty roughly proportional to the energy
$m=\mpl/\sqrt{N}$, that is
\be
\frac{\Delta\rh}{\expec{\hat r_{\rm H}}}
=
\frac{\sqrt{\expec{\hat r_{\rm H}^2-\expec{\hat r_{\rm H}}^2}}}{\expec{\hat r_{\rm H}}}
\simeq
\frac{1}{N}
\ ,
\ee
which vanishes as fast as in the previous case for large $N$, again as one
expects in a proper semiclassical regime~\cite{DvaliGomez}.
\par
The numerical analysis of Eq.~\eqref{CE1} displayed in Appendix~\ref{numCE}
shows the actual peak of the spectral function $C=C(E)$ is at slightly larger
values of $E$, and that the width is narrower, than the ones given
by the analytical approximation~\eqref{CE>M} when $N\gg 1$.
However, the uncertainties $\Delta\rh$ obtained so far, are very likely just lower bounds.
As we show in Appendix~\ref{mixS}, the spectral coefficient $C=C(E>M)$ contains 
$N$ contributions, displayed in Eq.~\eqref{CE3}, of which we have just tried to
include one (at a time) here. 
In particular, for $\gamma\simeq 1$, one could argue that all of the $N$ terms
in Eq.~\eqref{CE3} are relevant and their sum might yield a larger uncertainty.
If each of these terms contributes an uncertainty $\Delta\rh/\expec{\hat r_{\rm H}}\sim N^{-1}$,
like the terms already estimated, they could possibly add up to
$\Delta\rh\sim \expec{\hat r_{\rm H}}$~\footnote{We note in passing this is the
uncertainty one would obtain for a single quantum mechanical particle of mass
$m\gg\mpl$~\cite{Cfuzzy,GUPfuzzy}.
A numerical estimate of all of the $N$ terms in Eq.~\eqref{CE3} is in progress.}. 
This would signal the causal structure of the system is far from being classical.
\par
Let us stress again that cases with $\gamma\not\ll 1$ cannot be used to model
a BEC black hole, since then most or all of the scalars are in some excited mode
with $k>k_{\rm c}$.
However, it is not unreasonable to conjecture that these states play a role either at the
threshold of black hole formation (before the gravitons condense into the ground state
$\ket{k_{\rm c}}$) or near the end of black hole evaporation (when the black hole is
the hottest).
We shall further comment about this in the concluding Section.
\subsection{Black hole with thermal hair}
\label{ThBH}
We have seen in Section~\ref{BHhair} that, for $\gamma\ll 1$, the dominant contribution
to the spectral decomposition of the quantum state of $N$ scalars is given by the configuration 
with just one boson in the continuum of excited states, and the remaining $N-1$ in
the ground state.
In that Section, we employed a rather {\em ad hoc\/} Gaussian distribution for the continuous
part of the spectrum, but we then found the spectral function has a typical width of the order
of the Hawking temperature,
\be
T_{\rm H}
=
\frac{\mpl^2}{4\,\pi\,M}
\simeq
\frac{\mpl}{\sqrt{N}}
\ ,
\ee
or $T_{\rm H}\simeq m$.
\par
Let us therefore see what happens if we replace the Gaussian distribution
in Eq.~\eqref{psi_i} with a thermal spectrum at the temperature $T_{\rm H}$,
which is predicted according to the Hawking effect, that is
\be
\ket{\psi_{\rm S}^{(i)}}
\simeq
\mathcal{N}_\gamma
\left(
\ket{m}
+
\gamma\,\frac{e^{T_{\rm H}^{-1}\,\frac{m}{2}}}{\sqrt{T_{\rm H}}}
\int_{m}^\infty
\d E_i\,
e^{-T_{\rm H}^{-1}\,E_i}
\ket{E_i}
\right)
\ ,
\label{psiEth}
\ee
where the arbitrary coefficient $\gamma$ again weighs the relative probability 
of having a scalar quantum in the continuous spectrum with respect to it being
in the ground state.
In the same approximation $\gamma\ll 1$ as used in Section~\ref{BHhair},
the dominant correction to the ideal black hole is again given by the configuration
with just one boson in the continuum, for which
\be
C(E>M)
&\!\!\simeq\!\!&
\gamma
\left(\frac{e}{m}\right)^{1/2}
\int_{m}^\infty
\d E_1\,
e^{-\frac{E_1}{m}}
\,
\delta(E-M+m-E_1)
\nonumber
\\
&\!\!\simeq\!\!&
\gamma
\left(\frac{e}{m}\right)^{1/2}
e^{-\frac{E-(M-m)}{m}}
\ ,
\ee
where we used $T_{\rm H}\simeq m$.
Adding the contribution from the ground state, we obtain the (normalised)
non-vanishing spectral coefficients are given by
\be
C(E\ge M)
\simeq
\tilde{\mathcal{N}}_\gamma
\left[
\delta_{E,M}
+
\gamma
\left(\frac{e}{m}\right)^{1/2}
e^{-\frac{E-(M-m)}{m}}
\right]
\ ,
\ee
where the new normalisation factor is given by
\be
\tilde{\mathcal{N}}_\gamma
=
\left(1+\frac{\gamma^2}{2\,e}\right)^{-1/2}
\ .
\ee
\par
It is now easy to compute the expectation value of the total energy,
for large $N$ and to leading order in $\gamma$,
\be
\expec{E}
\simeq
\sqrt{N}\,\mpl
\left(
1+\frac{\gamma^2}{4\,e\,N}
\right)
\ ,
\ee
and its uncertainty
\be
\Delta E
\simeq
\frac{\gamma\,\mpl}{2\,\sqrt{e\,N}}
\ .
\ee
For $N\gg 1$, the above expressions lead to 
\be
\frac{\Delta E}{\expec{E}}
\sim
\frac{\gamma}{2\,\sqrt{e}\,N}
\ .
\ee
Up to an irrelevant numerical factor, this is the same behaviour we
found in the previous two cases, and one therefore expects the uncertainty
in the horizon size
\be
\frac{\Delta\rh}{\expec{\hat r_{\rm H}}}
\sim
\frac{1}{N}
\ ,
\ee
which, for large $N$, is the same decreasing behaviour we found previously
in Section~\ref{BHhair} for a Gaussian distribution~\cite{DvaliGomez}.
\par
This result was expected, since the thermal distribution in Eq.~\eqref{psiEth}
lends the excited boson a probability to have an energy $E_i>m$
comparable to that of the Gaussian distribution in Eq.~\eqref{psiEi},
which was specifically defined with the same width $\Delta_i=m=T_{\rm H}$. 
Correspondingly, the uncertainty $\Delta\rh\sim N^{-1/2}$ is approximately
the same.
\section{Concluding speculations}
\setcounter{equation}{0}
\label{conc}
In this work, we have considered a corpuscular model of black holes of
the form conjectured in Refs.~\cite{DvaliGomez}, and investigated its
causal structure by means of a formalism for quantum systems
we had previously developed in Refs.~\cite{Cfuzzy,GUPfuzzy,qhoop}.
We thus found the presence of an horizon whose size is in agreement
with the classical picture of a Schwarzschild black hole for large $N$
(when the energy of each scalar is much smaller than $\mpl$,
but the total energy is well above $\mpl$).
We also found the uncertainty in the horizon's size is typically of the order
of the energy of the expected Hawking quanta (albeit, for a suitably
chosen variety of states), the latter being proportional to $1/\sqrt{N}$
like it was claimed in Refs.~\cite{DvaliGomez}.
This result is in striking contrast with Eq.~\eqref{Mmp} for a single
particle of mass $m\gg\mpl$, for which a proper semiclassical
behaviour cannot be recovered, and thus further supports the conjecture
that black holes must be composite objects made of very light constituents.
\par
Based on the above picture, one may argue about what could be going
on during the gravitational collapse of a star.
In this respect, the case considered in Sections~\ref{Qcoe} represents
a simplistic model of a Newtonian lump of ordinary matter:
the source $J\sim\rho$ represents the star, with
$\Bra{g}\hat {\phi}\Ket{g}\simeq\phi_{\rm c}\sim V_{\rm N}$
that reproduces the outer Newtonian potential it generates.
In this situation, the energy contribution of the gravitons themselves
is negligible.
The cases analysed in Section~\ref{BH} are then just the opposite,
since the contribution of any matter source is neglected therein,
the quantum state is self-sustained, roughly confined
inside a region of size given by its Schwarzschild radius,
and (almost) monochromatic (see Section~\ref{BHstar}).
This state also satisfies the scaling relations~\eqref{mMax} and \eqref{MMax},
which have been known to hold for a self-gravitating body near
the threshold of black hole formation well before Ref.~\cite{DvaliGomez}
(see, for example, Refs.~\cite{Ruffini:1969qy,Chavanis:2011cz}).
In our treatment, all time dependence is frozen, and no connection
between the two configurations can be made explicitly.
However, one may view these two regimes as approximate descriptions 
of the beginning and (a possible) end-state of the collapse of a
proper star, as we briefly suggest in Section~\ref{BHstar}.
\par
If the above picture is to make sense, there might occur a phase transition
for the graviton state around the time when matter and gravitons have
comparable weight, much like what happens in a conductor
on the verge of superconductivity.
The analysis of this transition might be crucial in order to determine
whether the star indeed ever reaches the state of BEC black hole
(according to Refs.~\cite{DvaliGomez,flassig}, a black hole is precisely
the state at the quantum phase transition),
or just approaches that asymptotically, or, for any reasons, avoids it.
A starting point for tackling this issue in a toy model could be the
Klein-Gordon equation with both matter and ``graviton'' currents,
\be
\Box\phi(x)
=
q_{\rm M}\,J_{\rm M}(x)
+q_{\rm G}\,J_{\rm G}(x)
\ ,
\label{eqTot}
\ee
where $J_{\rm M}$ is the general matter current employed in Section~\ref{Qcoe}
and $J_{\rm G}$ the graviton current given in Eq.~\eqref{Jg}.
For the configuration corresponding to a star, one could treat the latter
as a perturbation, by formally expanding for small $q_{\rm G}$.
We can expect this approximation will lead to a correction for the Newtonian
potential at short distance from the star, since the graviton current~\eqref{Jg}
is formally equivalent to a mass term for the ``graviton'' $\phi$.
In the opposite regime, ordinary matter becomes a small correction,
and one could instead expand for small $q_{\rm M}$.
Outside the black hole, such a correction should then increase
the fuzziness of the horizon.
Of course, if a phase transition happens, it might be when neither sources
are negligible, so that perturbative arguments would fail to capture its
main features.
In any case, an {\em order parameter\/} should be identified.
\par
In fact, {\em close to the phase transition\/}, and before the black hole forms,
one might speculate that the case discussed in Section~\ref{noBH},
with $\gamma\gtrsim 1$ (in which all of the bosons are in a slightly excited
state above the BEC energy) hints at what physical processes could occur
thereabout.
We recall that the parameter $\gamma$ is essentially the relative probability of
finding each one of the $N$ bosons in an excited state with $k>k_{\rm c}$
rather than in the ground state with $k=k_{\rm c}$. 
Since the case $\gamma\ll 1$ of Section~\ref{BHhair} or, perhaps more
accurately, the largely equivalent Hawking case of Section~\ref{ThBH},
should represent the quantum state of a formed black hole,
one might conjecture that it is the parameter $\gamma$ (or a function
thereof) that can be viewed as an {\em effective\/} order parameter for the
transition from star to black hole.
In a truly dynamical context, $\gamma$ should furthermore acquire
a time dependence, thus decreasing from values of order one or larger
to much smaller figures along the collapse.
It hence appears worth investigating the possible dependence of $\gamma$
on the physical variables usually employed to define the state of matter
along the collapse, in order to identify a {\em physical\/} order parameter,
although this task will likely be much more difficult.
\par
Let us conclude by stressing the fact we have worked in the Newtonian
approximation for the special relativistic scalar equation, which basically
means we have relied on solutions of the Poisson equation in order
to describe the quantum states of gravity.
However, an appropriate description of black holes would naturally call
for general relativity.
In this respect, the only general relativistic aspect we have included is
the condition for the existence of trapping surfaces, as follows from the
Einstein equations for spherically symmetric systems, and which lies at the
foundations of the classical hoop conjecture, as well as of our horizon
wave-function (and the Generalized Uncertainty Principle that follows
from it~\cite{GUPfuzzy}).
It is important to remark that similarly ``fuzzy'' descriptions
of a black hole's horizon were recently derived from the quantisation
of spherically symmetric space-time metrics, which do not require any knowledge
of the quantum state of the source (see~\cite{davidson} and~\cite{ramy}
and references therein).
Other investigations of simple collapsing systems, such as thin
shells~\cite{torres} or thick shells~\cite{brustein}, also seem to
point towards similar scenarios.
We emphasize that our approach is much more general in that it allows one
to uniquely relate the causal structure of space-time, encoded by the
horizon wave-function $\psi_{\rm H}$, to the presence of any material source
in the state $\psi_{\rm S}$.
It is cleat that, in order to study the time evolution of the system,
a ``feedback'' from $\psi_{\rm H}$ into $\psi_{\rm S}$ must be introduced.
These deeply conceptual issues are left for future investigations.
\section*{Acknowledgements}
This work was supported in part by the European Cooperation
in Science and Technology (COST) action MP0905 ``Black Holes in a Violent  Universe''.
The work of O.~M.~is supported by UEFISCDI grant PN-II-RU-TE-2011-3-0184. 
\appendix
\section{$N$-boson spectrum}
\label{mixS}
\setcounter{equation}{0}
Starting from the single particle wave-functions~\eqref{psiEi},
the total wave-function of a system of $N$ bosons will be given
by the totally symmetrised product
\be
\ket{\psi_{\rm S}}
\simeq
\frac{1}{N!}\,
\sum_{\{\sigma_i\}}^N
\left[
\bigotimes_{i=1}^N
\,
\ket{\psi_{\rm S}^{(i)}}
\right]
=
\frac{1}{N!}\,
\sum_{\{\sigma_i\}}^N
\left[
\bigotimes_{i=1}^N
\left(
\ket{m}
+
\gamma\,
\int_{m}^\infty
\frac{\sqrt{2}\,\d E_i}{\sqrt{m\,\sqrt{\pi}}}\,
e^{-\frac{(E_i-m)^2}{2\,m^2}}
\ket{E_i}
\right)
\right]
\ ,
\label{NGaus}
\ee
where, here and in the following equations, $\sum_{\{\sigma_i\}}^N$ always denotes
the sum over all of the $N!$ permutations $\{\sigma_i\}$ of the $N$ terms inside
the square brackets, and we omit (irrelevant) overall normalisation factors of
$\mathcal{N}_\gamma$ for the sake of simplicity.
Note that we can group equal powers of $\gamma$ in the above expression and obtain
\be
\ket{\psi_{\rm S}}
&\!\!\simeq\!\!&
\frac{1}{N!}\,
\sum_{\{\sigma_i\}}^N
\left[
\bigotimes_{i=1}^N
\ket{m}
\right]
\nonumber
\\
&&
+
\frac{\gamma}{N!}
\left(\frac{2}{m\,\sqrt{\pi}}\right)^{1/2}
\,
\sum_{\{\sigma_i\}}^{N}
\left[
\bigotimes_{i=2}^{N}
\ket{m}
\otimes
\int_{m}^\infty
\d E_1\,
e^{-\frac{(E_1-m)^2}{2\,m^2}}
\ket{E_1}
\right]
\nonumber
\\
&&
+
\frac{\gamma^2}{N!}
\left(\frac{2}{m\,\sqrt{\pi}}\right)
\sum_{\{\sigma_i\}}^{N}
\left[
\bigotimes_{i=3}^{N}
\ket{m}
\otimes
\int_{m}^\infty
\d E_1\,
e^{-\frac{(E_1-m)^2}{2\,m^2}}
\ket{E_1}
\otimes
\int_{m}^\infty
\d E_2\,
e^{-\frac{(E_2-m)^2}{2\,m^2}}
\ket{E_2}
\right]
\nonumber
\\
&&
+
\ldots
\nonumber
\\
&&
+
\frac{\gamma^J}{N!}
\left(\frac{2}{m\,\sqrt{\pi}}\right)^{J/2}\,
\sum_{\{\sigma_i\}}^{N}
\left[
\bigotimes_{i=J+1}^{N}
\ket{m}
\,
\bigotimes_{j=1}^{J}
\int_{m}^\infty
\d E_j\,
e^{-\frac{(E_j-m)^2}{2\,m^2}}
\ket{E_j}
\right]
\nonumber
\\
&&
+
\ldots
\nonumber
\\
&&
+
\frac{\gamma^N}{N!}
\left(\frac{2}{m\,\sqrt{\pi}}\right)^{N/2}\,
\sum_{\{\sigma_i\}}^{N}
\left[
\bigotimes_{i=1}^{N}
\int_{m}^\infty
\d E_i\,
e^{-\frac{(E_i-m)^2}{2\,m^2}}
\ket{E_i}
\right]
\ ,
\label{NGausG}
\ee
where the power of $\gamma$ clearly equals the number of bosons in a
continuous (``excited'') mode with $k>k_{\rm c}$.
This form will help in obtaining the spectral decomposition.
\par
In fact, one of course finds $C(E<M)=0$, and
\be
C(M)
\simeq
\bra{M}
\frac{1}{N!}
\sum_{\{\sigma_i\}}^N
\left[
\bigotimes_{i=1}^N
\ket{m}
\right]
=
N!
\bra{M}
\frac{1}{N!}\,
\bigotimes_{i=1}^N
\ket{m}
=
1
\ ,
\label{CE=M}
\ee
since the energy $E$ can equal $M=N\,m$ only when all of the $N$
bosons are in the ground state of lowest momentum number
$k_{\rm c}=m/\hbar$.
For $E>M$, the term of order $\gamma^0$ instead does not contribute
and one obtains
\be
C(E>M)
&\!\!\simeq\!\!&
\gamma
\left(\frac{2}{m\,\sqrt{\pi}}\right)^{1/2}
\int_{m}^\infty
\d E_1\,
e^{-\frac{(E_1-m)^2}{2\,m^2}}
\,
\delta(E-M+m-E_1)
\nonumber
\\
&&
+
\gamma^2
\left(\frac{2}{m\,\sqrt{\pi}}\right)
\int_{m}^\infty
\d E_1\,
\int_{m}^\infty
\d E_2\,
e^{-\frac{(E_1-m)^2}{2\,m^2}-\frac{(E_2-m)^2}{2\,m^2}}
\,
\delta(E-M+2\,m-E_1-E_2)
\nonumber
\\
&&
+
\ldots
\nonumber
\\
&&
+
\gamma^J
\left(\frac{2}{m\,\sqrt{\pi}}\right)^{J/2}
\int_{m}^\infty
\d E_1\cdots
\int_{m}^\infty
\d E_J\,
\exp\left\{-\sum_{j=1}^J\frac{(E_j-m)^2}{2\,m^2}\right\}
\nonumber
\\
&&
\phantom{\gamma^J
\left(\frac{2}{m\,\sqrt{\pi}}\right)^{J/2}
\int_{m}^\infty}
\times
\delta\left(E-M+J\,m-\sum_{j=1}^J E_j\right)
\nonumber
\\
&&
+
\ldots
\nonumber
\\
&&
+
\gamma^N
\left(\frac{2}{m\,\sqrt{\pi}}\right)^{N/2}
\int_m^\infty
\d E_1\cdots
\int_m^\infty
\d E_N\,
\exp\left\{-\sum_{i=1}^N\frac{(E_i-m)^2}{2\,m^2}\right\}
\nonumber
\\
&&
\phantom{\gamma^J
\left(\frac{2}{m\,\sqrt{\pi}}\right)^{J/2}
\int_{m}^\infty}
\times
\delta\left(E-\sum_{i=1}^N E_i\right)
\ ,
\ee
which can be further simplified to
\be
C(E>M)
&\!\!\simeq\!\!&
\gamma
\left(\frac{2}{m\,\sqrt{\pi}}\right)^{1/2}
e^{-\frac{(E-M)^2}{2\,m^2}}
\nonumber
\\
&&
+
\gamma^2
\left(\frac{2}{m\,\sqrt{\pi}}\right)
\int_{m}^\infty
\d E_1\,
\exp\left\{-\frac{(E_1-m)^2}{2\,m^2}-\frac{(E-M+m-E_1)^2}{2\,m^2}\right\}
\nonumber
\\
&&
+
\ldots
\nonumber
\\
&&
+
\gamma^N
\left(\frac{2}{m\,\sqrt{\pi}}\right)^{N/2}
\int_m^\infty
\d E_1\cdots
\int_m^\infty
\d E_N\,
\exp\left\{-\sum_{i=1}^N\frac{(E_i-m)^2}{2\,m^2}\right\}
\nonumber
\\
&&
\phantom{\gamma^J
\left(\frac{2}{m\,\sqrt{\pi}}\right)^{J/2}
\int_{m}^\infty}
\times
\,\delta\left(E-\sum_{i=1}^N E_i\right)
\ .
\label{CE3}
\ee
For fixed $m$ (or $N$) and $\gamma\ll 1$, one can just keep the first term
in Eq.~\eqref{CE3} above (this contribution is analysed in details in
Section~\ref{BHhair}).
Conversely, for $\gamma\gtrsim 1$, it is the last term that dominates,
which is estimated analytically in Appendix~\ref{contCE} and leads
to the results presented in Section~\ref{noBH}.
For a numerical calculation of this spectral coefficient~\eqref{CE1},
see also Appendix~\ref{numCE}.
\par
We end this Appendix with a word of caution.
For $\gamma\simeq 1$, all terms could equally contribute, and their
(cumbersome) evaluation is left for future investigations. 
However, we can already note here that, upon considering $m\simeq\mpl/\sqrt{N}$,
one can identify the alternative expansion parameter $\tilde\gamma=\gamma^4\,N$
in front of each contribution in Eq.~\eqref{CE3} above.
The first term, of order $\tilde\gamma^{1/4}$, would hence seem to dominate
for $\tilde\gamma\ll 1$, or $\gamma\ll N^{-1/4}$,
which means a very tight bound on $\gamma$ for macroscopic black holes.
And the last term in Eq.~\eqref{CE3} should likewise dominate for
$\gamma\gtrsim N^{-1/4}$.
This remark makes it clear that the interplay between the small $\gamma$ 
expansion and the large $N$ expansion is not trivial, and the validity of the
final results can safely be assessed only by checking {\em a posteriori\/}
that higher-order terms (in each expansion) are smaller than lower-order terms.
Indeed, this condition holds true for the cases studied in the main text,
providing one expands in $\gamma$ first and then takes $N$ large.
\section{Analytical spectrum for $\gamma\gtrsim 1$}
\label{contCE}
\setcounter{equation}{0}
We start by noting the spectral coefficient in Eq.~\eqref{CE1} can be written as
\be
C(E\ge M)
&\!\!\sim\!\!&
\int_m^\infty
\d E_1\cdots
\int_m^\infty
\d E_N\,
\exp\left\{-\sum_{i=1}^N\frac{(E_i-m)^2}{2\,m^2}\right\}
\,\delta\left(E-\sum_{i=1}^N E_i\right)
\nonumber
\\
&\!\!\sim\!\!&
\int_m^\infty
\d E_1\cdots
\int_m^\infty
\d E_{N-1}\,
\exp\left\{-\sum_{i=1}^{N-1}\frac{(E_i-m)^2}{2\,m^2}
-\frac{\left(E-\sum_{i=1}^{N-1}E_i-m\right)^2}{2\,m^2}
\right\}
\nonumber
\\
&\!\!\equiv\!\!&
\int_m^\infty
\d E_1\cdots
\int_m^\infty
\d E_{N-1}
\,e^{-\frac{F^2(E,\{\mathcal{E}_i\})}{2\,m^2}}
\ .
\label{CE2}
\ee
In order to proceed, we find it convenient to write the argument
of the exponential as
\be
-2\,m^2\,F^2(E,\{\mathcal{E}_i\})
&\!\!\equiv\!\!&
\sum_{i=1}^{N-1} (E_i-m)^2
+
\left(E-\sum_{i=1}^{N-1}E_i-m\right)^2
\nonumber
\\
&\!\!=\!\!&
\sum_{i=1}^{N-1} (E_i-m)^2
+
\left[E-\sum_{i=1}^{N-1}(E_i-m)-N\,m\right]^2
\nonumber
\\
&\!\!=\!\!&
\left(E-M\right)^2
+
\sum_{i=1}^{N-1}\mathcal{E}_i^2
+
\left[
\sum_{i=1}^{N-1}\mathcal{E}_i
-2\,(E-M)
\right]
\sum_{j=1}^{N-1}\mathcal{E}_j
\ ,
\ee
where $\mathcal{E}_i=E_i-m$, so that
\be
C(E\ge M)
&\!\!\sim\!\!&
e^{-\frac{(E-M)^2}{2\,m^2}}
\int_0^\infty
\d \mathcal{E}_1\cdots
\int_0^\infty\d \mathcal{E}_{N-1}
\nonumber
\\
&&
\times
\exp\left\{-\sum_{i=1}^{N-1}\frac{\mathcal{E}_i^2}{2\,m^2}
-
\left[
\sum_{i=1}^{N-1}\frac{\mathcal{E}_i}{2\,m}
-\frac{(E-M)}{m}
\right]
\sum_{j=1}^{N-1} \frac{\mathcal{E}_j}{m}
\right\}
\nonumber
\\
&\!\!\equiv\!\!&
e^{-\frac{(E-M)^2}{2\,m^2}}
\,
I(E,M;m)
\ .
\label{B3}
\ee
We then note that the above integral contains the Gaussian measure, 
\be
\int_0^\infty
\d \mathcal{E}_1\cdots
\int_0^\infty\d \mathcal{E}_{N-1}
\exp\left\{-\sum_{i=1}^{N-1}\frac{\mathcal{E}_i^2}{2\,m^2}\right\}
\sim
\int_0^\infty
\mathcal{E}^{N-2}\,\d \mathcal{E}
\exp\left\{-\frac{\mathcal{E}^2}{2\,m^2}\right\}
\ ,
\ee
where $\mathcal{E}^2=\sum_{i=1}^{N-1}{\mathcal{E}_i^2}$,
and is significantly different from zero only for $\mathcal{E}\lesssim m$.
We can therefore approximate 
\be
\sum_{i=1}^{N-1}{\mathcal{E}_i}
\simeq
\mathcal{E}
\ ,
\ee
from which we obtain
\be
I(E,M;m)
&\!\!\sim\!\!&
\int_0^\infty
\mathcal{E}^{N-2}\,\d \mathcal{E}
\exp\left\{-\frac{\mathcal{E}^2}{m^2}
+\frac{(E-M)}{m}\,\frac{\mathcal{E}}{m}
\right\}
\nonumber
\\
&\!\!=\!\!&
e^{\frac{(E-M)^2}{4\,m^2}}
\int_0^\infty
\mathcal{E}^{N-2}\,\d \mathcal{E}
\exp\left\{-\frac{\left[2\,\mathcal{E}-(E-M) \right]^2}{4\,m^2}
\right\}
\ .
\ee
This integral can be exactly evaluated in terms of hypergeometric functions,
but we can just further approximate it here as
\be
I(E,M;m)
&\!\!\sim\!\!&
m\,(E-M)\,P_{(N-3)}(E,M)\,
e^{\frac{(E-M)^2}{4\,m^2}}
\sim
(E-M)\,
e^{\frac{(E-M)^2}{4\,m^2}}
\ ,
\ee
where $P_{(N-3)}$ is a polynomial of degree $N-3$.
For $N=M/m\gg1$, we finally obtain,
\be
C(E\ge M)
&\!\!\sim\!\!&
(E-M)\,
e^{-\frac{(E-M)^2}{4\,m^2}}
\ ,
\ee
which is the expression in Eq.~\eqref{CE>M}.
\section{Numerical spectrum for $\gamma\gtrsim 1$}
\label{numCE}
\setcounter{equation}{0}
In this Appendix, we estimate the spectral coefficient in Eq.~\eqref{B3}
[which exactly equals the one in Eq.~\eqref{CE1}] for various values of $N$.
For this purpose, we have implemented a standard Monte Carlo method
in a \textsc{Mathematica\/} notebook, in which the coefficient is also
numerically normalised, so that 
\be
\int_M^\infty
C^2(E)\,\d E
= 1
\ .
\label{Nce}
\ee
The dependence of the spectral coefficient on the total energy $E$
is then compared with the analytical approximation~\eqref{CE>M}.
\par
\begin{figure}[h]
\centering
\includegraphics[width=8cm]{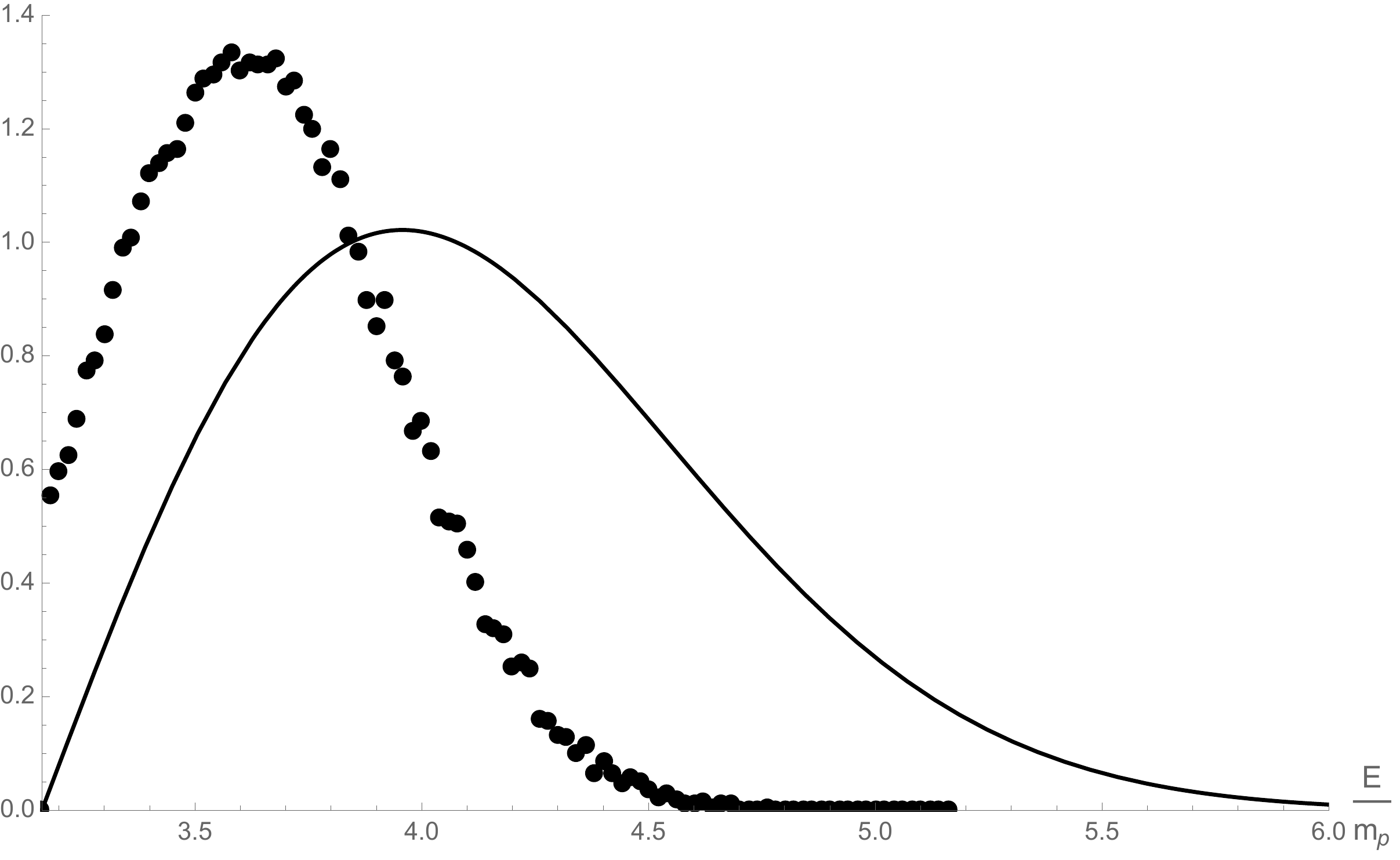}
$\ $
\includegraphics[width=8cm]{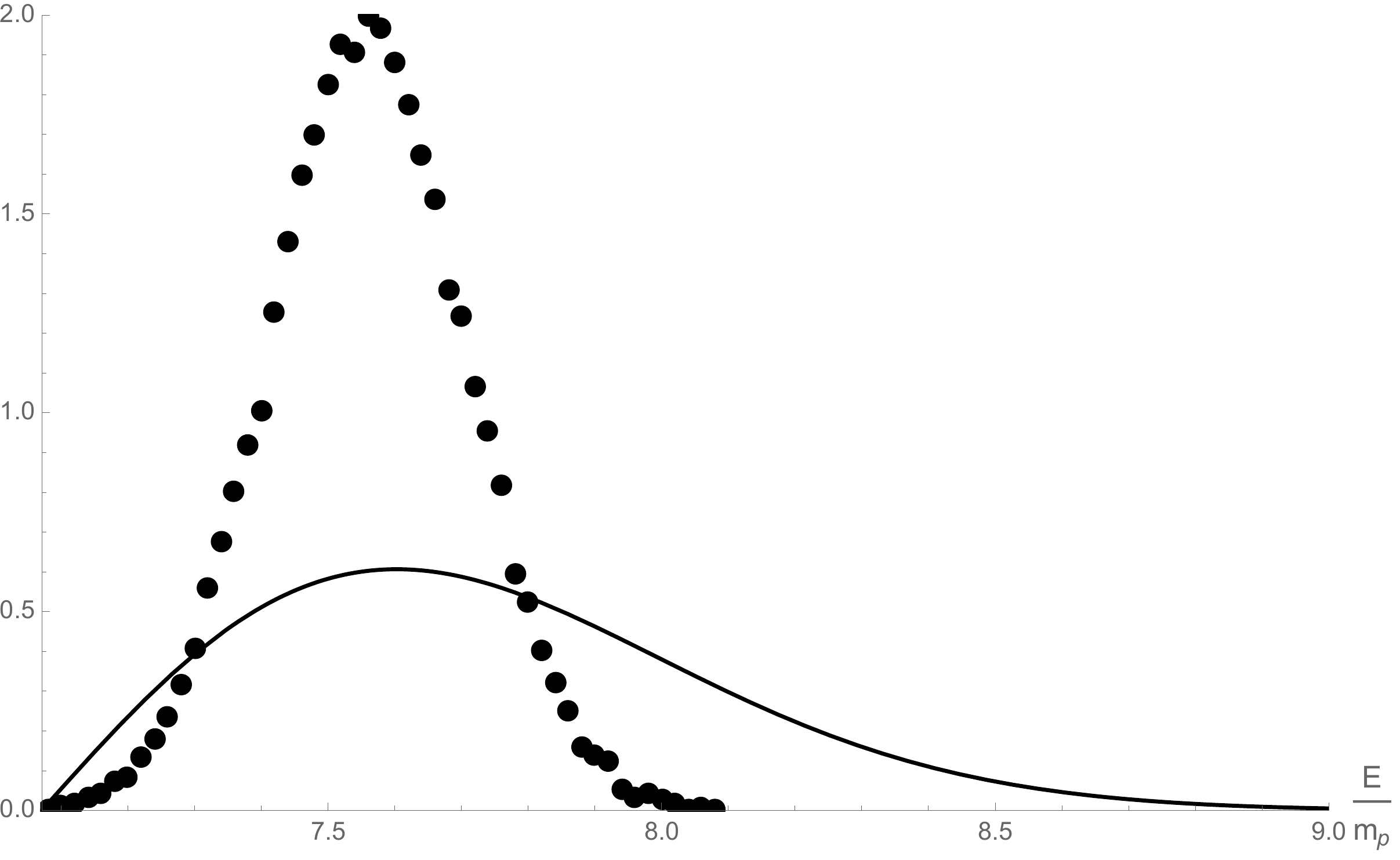}
\caption{Monte Carlo estimate of the spectral coefficient in Eq.~\eqref{B3} (dots)
compared to its analytical approximation~\eqref{CE>M} (solid line),
both normalised according to Eq.~\eqref{Nce},
for $N=10$ (left panel) and $N=50$ (right panel).
\label{CN10}}
\end{figure}
Fig.~\ref{CN10} shows this comparison for $N=10$ and $N=50$.
For the former value, the analytical approximation overestimates both the location of
the peak and (slightly) the width of the curve (thus underestimating the height of the peak).
For $N=50$, the location of the peak is instead very well identified by Eq.~\eqref{CE>M}, 
but the actual width remains narrower than its analytical approximation (resulting
in a large discrepancy in the peak values).
\par
\begin{figure}[h]
\centering
\includegraphics[width=8cm]{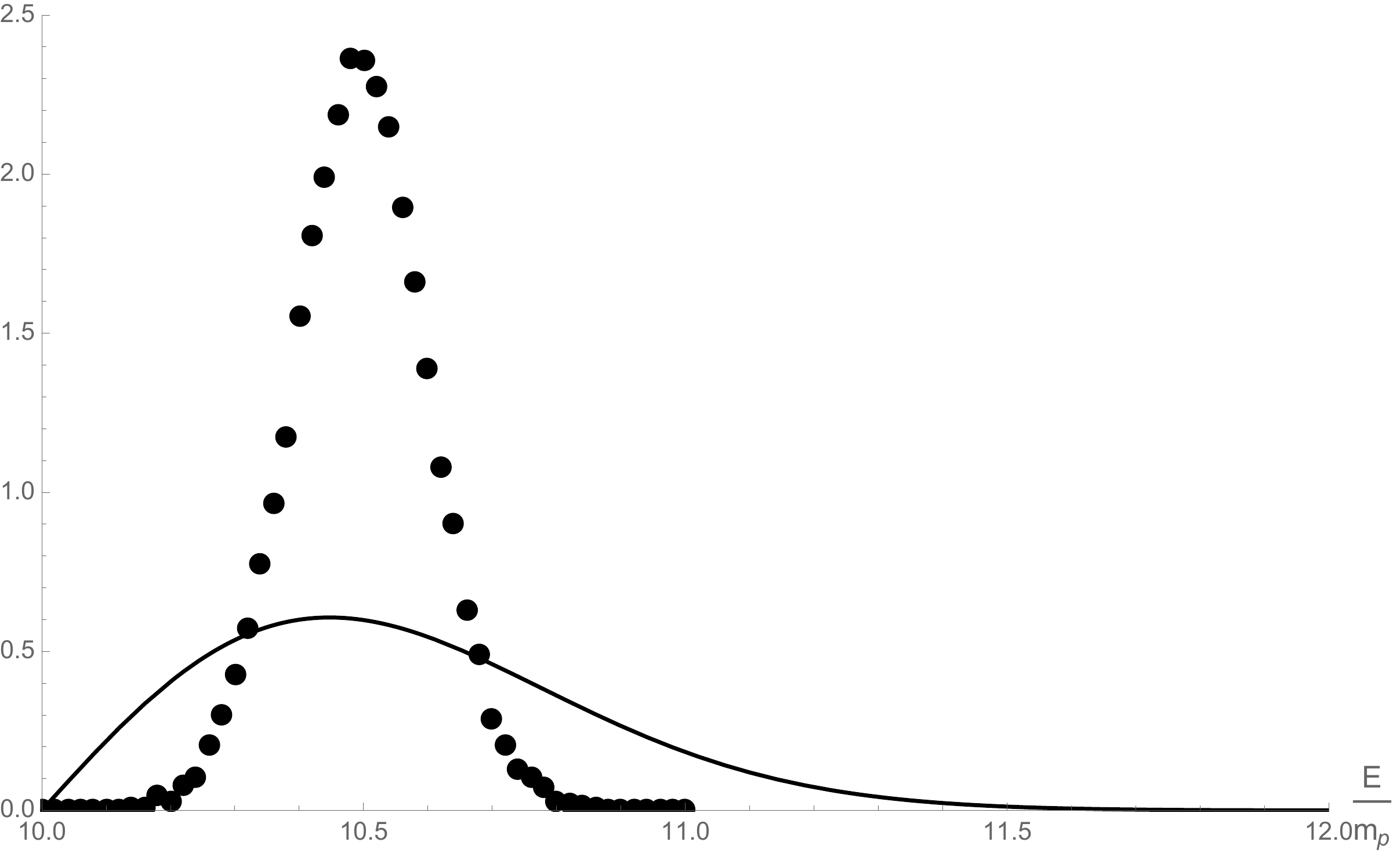}
$\ $
\includegraphics[width=8cm]{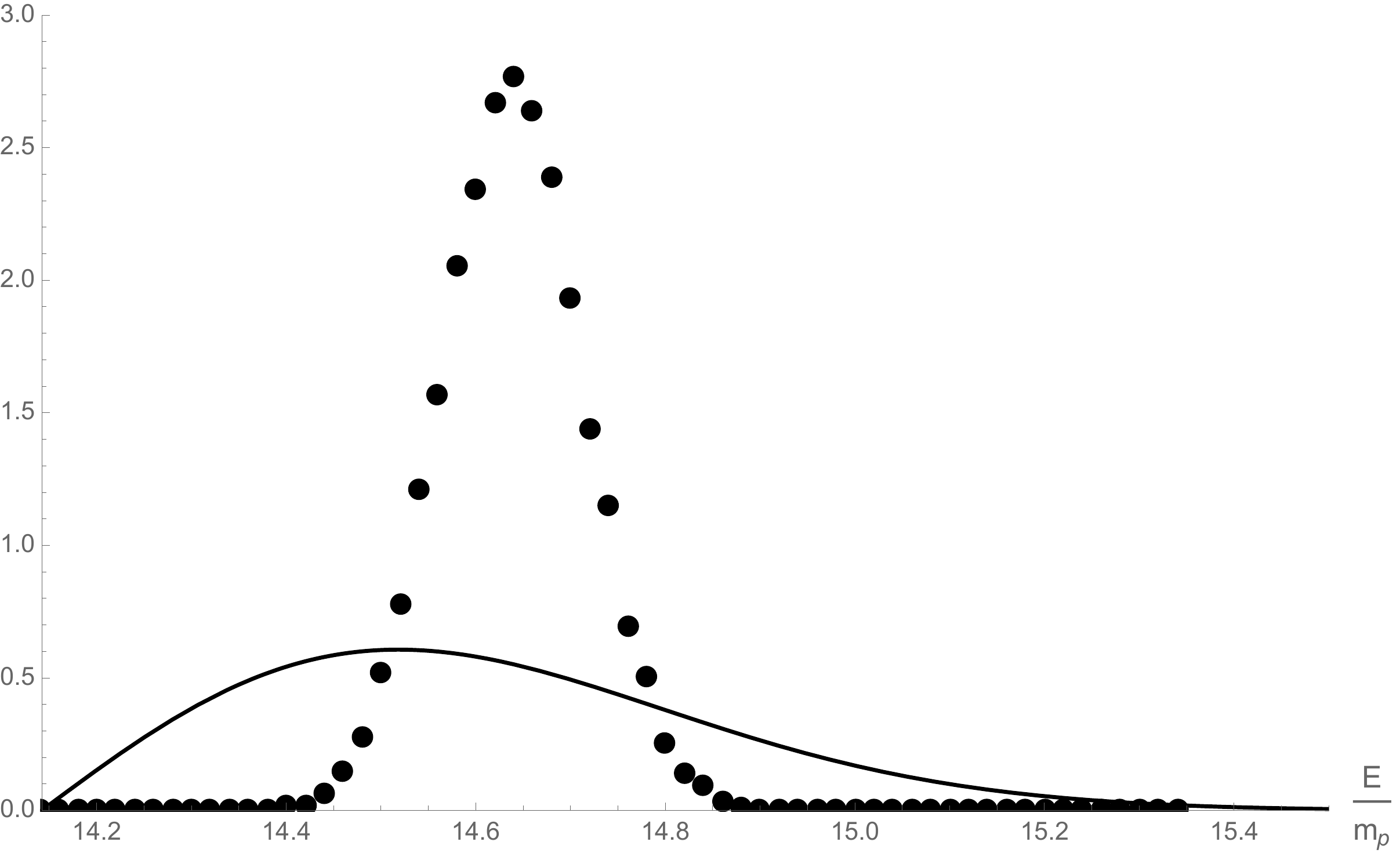}
\caption{Monte Carlo estimate of the spectral coefficient in Eq.~\eqref{B3} (dots)
compared to its analytical approximation~\eqref{CE>M} (solid line),
both normalised according to Eq.~\eqref{Nce},
for $N=100$ (left panel) and $N=200$ (right panel).
\label{CN100}}
\end{figure}
Fig.~\ref{CN100} shows the comparison for $N=100$ and $N=200$.
From these plots, it is clear that the analytical approximation progressively
underestimates the value of the energy at which the spectral coefficients peak,
at the same time overestimating more and more the width of the curve
(by about a factor of three in these two plots), and consequently underestimates
the height of the curve at peak values.
\par
\begin{figure}[h]
\centering
\includegraphics[width=8cm]{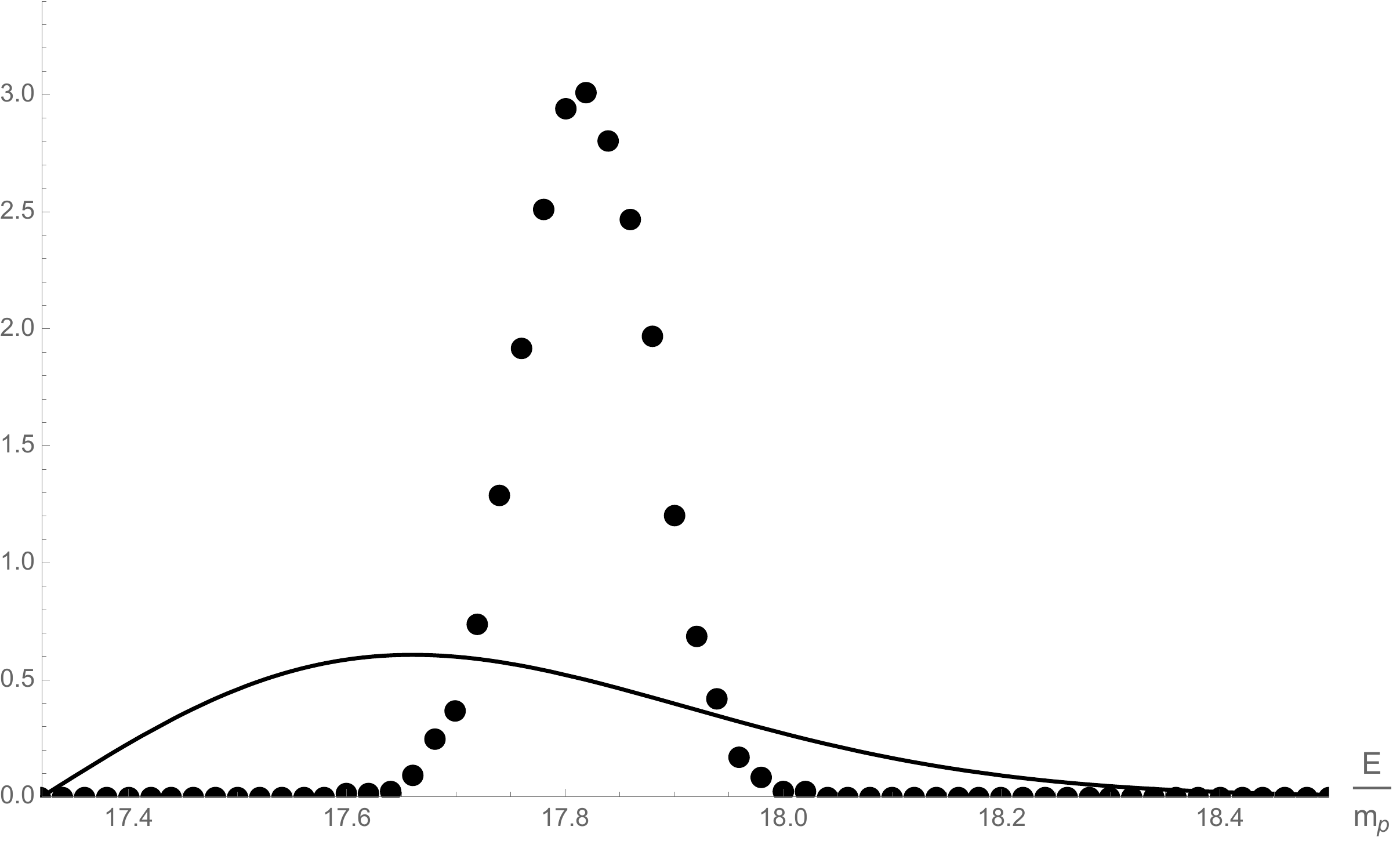}
$\ $
\includegraphics[width=8cm]{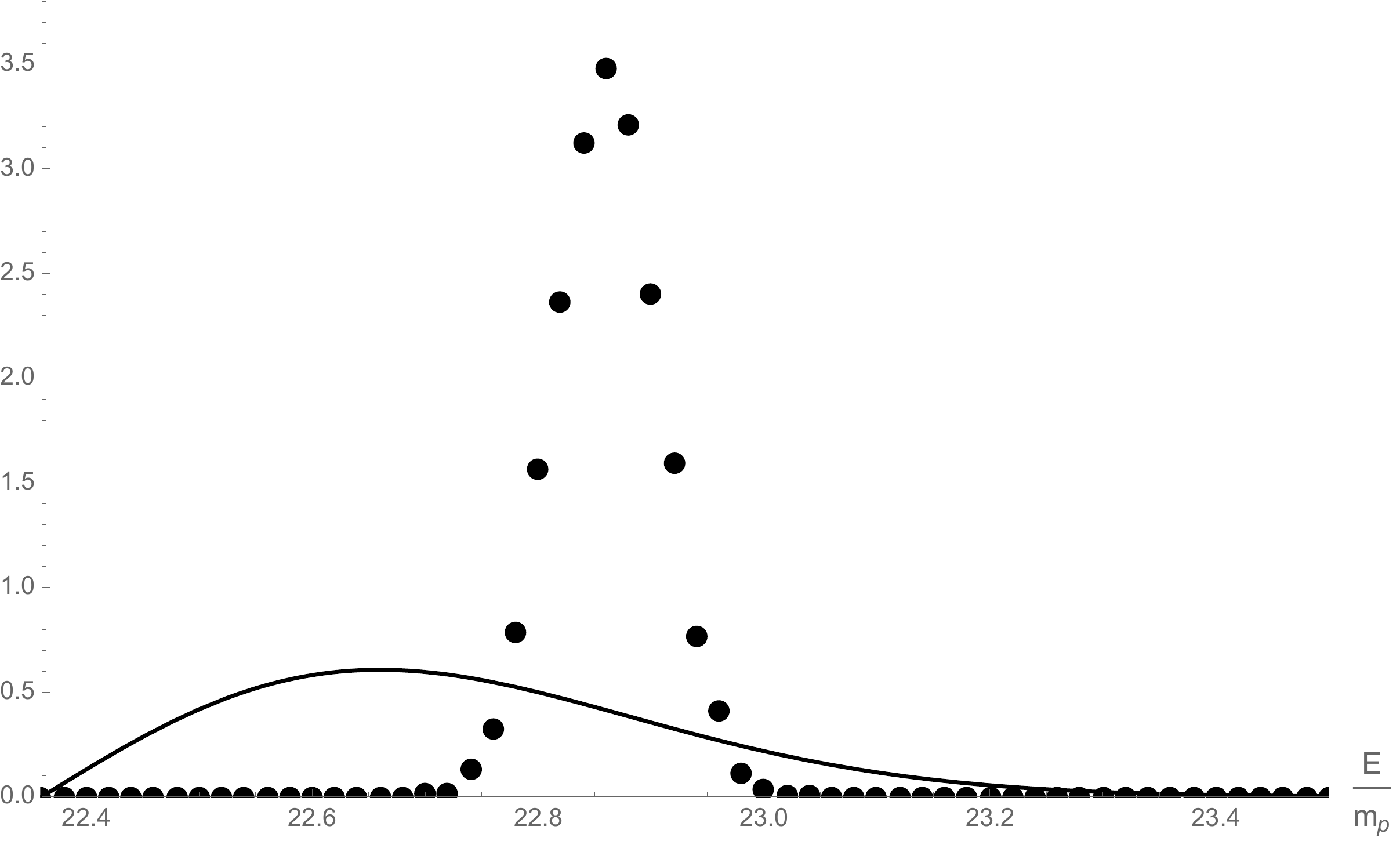}
\caption{Monte Carlo estimate of the spectral coefficient in Eq.~\eqref{B3} (dots)
compared to its analytical approximation~\eqref{CE>M} (solid line),
both normalised according to Eq.~\eqref{Nce},
for $N=300$ (left panel) and $N=500$ (right panel).
\label{CN500}}
\end{figure}
All of these trends are further confirmed in Fig.~\ref{CN500}, which shows
the comparison for $N=300$ and $N=500$.
The peaks of the numerical estimate and the analytical approximation
continue to separate further apart, whereas the numerical width becomes
narrower than the analytical width for increasing $N$.
\par
The overall conclusion is that the analytical approximation~\eqref{CE>M}
is fairly good for estimating the energy at the peak of the spectral coefficients,
but overestimates (underestimates) significantly the width (peak value),
for $N\gtrsim 100$.

\begin{thebibliography}{99}
%
%
\bibitem{DvaliGomez} 
G.~Dvali and C.~Gomez,
%``Quantum Compositeness of Gravity: Black Holes, AdS and Inflation'',
JCAP {\bf 01}, 023 (2014);
% arXiv:1312.4795 [hep-th];
``Black Hole's Information Group'',
arXiv:1307.7630;
%``Black Holes as Critical Point of Quantum Phase Transition,''
Eur.\ Phys.\ J.\ C {\bf 74}, 2752 (2014);
%[arXiv:1207.4059 [hep-th]];
%``Black Hole's 1/N Hair,''
Phys.\ Lett.\ B {\bf 719}, 419 (2013);
%[arXiv:1203.6575 [hep-th]];
%``Landau-Ginzburg Limit of Black Hole's Quantum Portrait: Self Similarity and Critical Exponent,''
Phys.\ Lett.\ B {\bf 716}, 240 (2012);
%[arXiv:1203.3372 [hep-th]];
%``Black Hole's Quantum N-Portrait,''
Fortsch.\ Phys.\  {\bf 61}, 742 (2013);
%[arXiv:1112.3359 [hep-th]];
G.~Dvali, C.~Gomez and S.~Mukhanov,
``Black Hole Masses are Quantized,''
arXiv:1106.5894 [hep-ph].
%
%
\bibitem{Kuhnel:2014oja}
F.~Kuhnel,
``Bose-Einstein Condensates with Derivative and Long-Range Interactions as Set-Ups for Analog Black Holes,''
arXiv:1312.2977 [gr-qc];
%%CITATION = ARXIV:1312.2977;%%
%1 citations counted in INSPIRE as of 21 Feb 2014
F.~Kuhnel and B.~Sundborg,
``Modified Bose-Einstein Condensate Black Holes in d Dimensions,''
arXiv:1401.6067 [hep-th];
%%CITATION = ARXIV:1401.6067;%%
%F.~Kuhnel and B.~Sundborg,
``High-Energy Gravitational Scattering and Bose-Einstein Condensates of Gravitons,''
arXiv:1406.4147 [hep-th].
%%CITATION = ARXIV:1406.4147;%%
%1 citations counted in INSPIRE as of 15 Sep 2014
%
%\cite{Kuhnel:2013uxa}
\bibitem{kuhnel2}
F.~Kuhnel and B.~Sundborg,
``Decay of Graviton Condensates and their Generalizations in Arbitrary Dimensions,''
arXiv:1405.2083 [hep-th].
%%CITATION = ARXIV:1405.2083;%%
%
%\cite{Mueck:2013wba}
\bibitem{Mueck:2013wba}
W.~Mueck,
%``Counting Photons in Static Electric and Magnetic Fields,''
Eur.\ Phys.\ J.\ C {\bf 73} (2013) 2679.
%[arXiv:1310.6909 [hep-th]].
%%CITATION = ARXIV:1310.6909;%%
%1 citations counted in INSPIRE as of 21 Feb 2014
%
%\cite{Casadio:2013hja}
\bibitem{Casadio:2013hja}
R.~Casadio and A.~Orlandi,
%``Quantum Harmonic Black Holes,''
JHEP {\bf 1308} (2013) 025.
%[arXiv:1302.7138 [hep-th]].
%%CITATION = ARXIV:1302.7138;%%
 %8 citations counted in INSPIRE as of 21 Feb 2014
%\cite{Berkhahn:2013woa}
%
\bibitem{Berkhahn:2013woa}
F.~Berkhahn, S.~Muller, F.~Niedermann and R.~Schneider,
%``Microscopic Picture of Non-Relativistic Classicalons,''
JCAP {\bf 1308} (2013) 028.
%[arXiv:1302.6581 [hep-th]].
%%CITATION = ARXIV:1302.6581;%%
%5 citations counted in INSPIRE as of 21 Feb 2014
%
%\cite{Flassig:2012re}
\bibitem{flassig}
D.~Flassig, A.~Pritzel and N.~Wintergerst,
%``Black Holes and Quantumness on Macroscopic Scales,''
Phys.\ Rev.\ D {\bf 87} (2013) 084007.
%[arXiv:1212.3344].
%%CITATION = ARXIV:1212.3344;%%
%9 citations counted in INSPIRE as of 21 Feb 2014
%
%\cite{Binetruy:2012kx}
\bibitem{Binetruy:2012kx}
P.~Binetruy,
``Vacuum energy, holography and a quantum portrait of the visible Universe,''
arXiv:1208.4645 [gr-qc].
%%CITATION = ARXIV:1208.4645;%%
%5 citations counted in INSPIRE as of 21 Feb 2014
%
%%%%%%SELF GRAVITATING STUFF%%%%%%%%%%%%
%\cite{Ruffini:1969qy}
%  
\bibitem{mueckPT}
W.~M\"uck and G.~Pozzo,
``Quantum Portrait of a Black Hole with P\"oschl-Teller Potential,''
arXiv:1403.1422 [hep-th].
%%CITATION = ARXIV:1403.1422;%% 
%  
\bibitem{bek}
J.~D.~Bekenstein,
%``Black holes and the second law,''
Lett.\ Nuovo Cim.\  {\bf 4} (1972) 737;
%%CITATION = NCLTA,4,737;%%
%586 citations counted in INSPIRE as of 06 Jun 2014
Phys.\ Rev.\ D {\bf 7} (1973) 2333.
%%CITATION = PHRVA,D7,2333;%%
%2601 citations counted in INSPIRE as of 06 Jun 2014
%
\bibitem{Cfuzzy} 
R.~Casadio,
``Localised particles and fuzzy horizons: A tool for probing Quantum Black Holes,''
arXiv:1305.3195 [gr-qc];
%%CITATION = ARXIV:1305.3195;%%
%3 citations counted in INSPIRE as of 10 Apr 2014
``What is the Schwarzschild radius of a quantum mechanical particle?,''
arXiv:1310.5452 [gr-qc].
 %%CITATION = ARXIV:1310.5452;%%
%
\bibitem{Ruffini:1969qy}
R.~Ruffini and S.~Bonazzola,
%``Systems of selfgravitating particles in general relativity and the concept of an equation of state,''
Phys.\ Rev.\  {\bf 187} (1969) 1767.
%%CITATION = PHRVA,187,1767;%%
%265 citations counted in INSPIRE as of 21 Feb 2014
%
%\cite{Colpi:1986ye}
\bibitem{Colpi:1986ye}
M.~Colpi, S.~L.~Shapiro and I.~Wasserman,
%``Boson Stars: Gravitational Equilibria of Selfinteracting Scalar Fields,''
 Phys.\ Rev.\ Lett.\  {\bf 57} (1986) 2485;
%%CITATION = PRLTA,57,2485;%%
%215 citations counted in INSPIRE as of 21 Feb 2014
%
%\cite{Membrado:1989ke}
%\bibitem{Membrado:1989ke}
  M.~Membrado, J.~Abad, A.~F.~Pacheco and J.~Sanudo,
  %``Newtonian Boson Spheres,''
  Phys.\ Rev.\ D {\bf 40} (1989) 2736;
  %%CITATION = PHRVA,D40,2736;%%
%\cite{Balakrishna:1999sv}
%
%\bibitem{Balakrishna:1999sv}
J.~Balakrishna,
``A Numerical study of boson stars: Einstein equations with a matter source,''
arXiv:gr-qc/9906110;
%%CITATION = GR-QC/9906110;%%
%8 citations counted in INSPIRE as of 21 Feb 2014
%\cite{Nieuwenhuizen:2008zz}
%
%\bibitem{Nieuwenhuizen:2008zz}
T.~M.~Nieuwenhuizen,
%``Supermassive Black Holes as Giant Bose-Einstein Condensates,''
Europhys.\ Lett.\  {\bf 83} (2008) 10008;
%[arXiv:0807.0315 [gr-qc]].
%%CITATION = ARXIV:0807.0315;%%
%5 citations counted in INSPIRE as of 21 Feb 2014
%
%\cite{Nieuwenhuizen:2009px}
%\bibitem{Nieuwenhuizen:2009px}
T.~M.~Nieuwenhuizen and V.~Spicka,
``Bose-Einstein condensed supermassive black holes:
A Case of renormalized quantum field theory in curved space-time,''
arXiv:0910.5377 [gr-qc];
%%CITATION = ARXIV:0910.5377;%%
%\cite{Chavanis:2011cz}
%
\bibitem{Chavanis:2011cz}
P.-H.~Chavanis and T.~Harko,
%``Bose-Einstein Condensate general relativistic stars,''
Phys.\ Rev.\ D {\bf 86} (2012) 064011.
%[arXiv:1108.3986 [astro-ph.SR]].
%%CITATION = ARXIV:1108.3986;%%
%17 citations counted in INSPIRE as of 21 Feb 2014
%
\bibitem{dvaliCL}
G.~Dvali and C.~Gomez,
``Self-Completeness of Einstein Gravity,''
arXiv:1005.3497 [hep-th];
%%CITATION = ARXIV:1005.3497;%%
%
G.~Dvali, G.F.~Giudice, C.~Gomez and A.~Kehagias,
%``UV-Completion by Classicalization,''
JHEP {\bf 1108} (2011) 108;
%[arXiv:1010.1415 [hep-ph]].
%%CITATION = ARXIV:1010.1415;%%
%
G.~Dvali, C.~Gomez and A.~Kehagias,
%``Classicalization of Gravitons and Goldstones,''
JHEP {\bf 1111}, 070 (2011);
%[arXiv:1103.5963 [hep-th]].
  %%CITATION = ARXIV:1103.5963;%%
  %34 citations counted in INSPIRE as of 15 Sep 2014%
G.~Dvali,  A.~Franca and C.~Gomez,
``Road signs to UV-completion,''
%JHEP {\bf 1108} (2011) 108.
arXiv:1204.6388 [hep-th].
%
\bibitem{GUPfuzzy} 
R.~Casadio and F.~Scardigli,
%``Horizon wave-function for single localized particles: GUP and quantum black hole decay,''
Eur.\ Phys.\ J.\ C {\bf 74}, 2685 (2014).
%[arXiv:1306.5298 [gr-qc]].
%%CITATION = ARXIV:1306.5298;%%
%5 citations counted in INSPIRE as of 10 Apr 2014
%
\bibitem{qhoop} 
R.~Casadio, O.~Micu and F.~Scardigli,
%``Quantum hoop conjecture: Black hole formation by particle collisions,''
Phys. Lett. {\bf B 732} (2014) 105.
%[arXiv:1311.5698 [hep-th]].
%%CITATION = ARXIV:1311.5698;%%
%2 citations counted in INSPIRE as of 10 Apr 2014
%
\bibitem{duff} 
M.~J.~Duff,
%``Quantum Tree Graphs and the Schwarzschild Solution,''
Phys.\ Rev.\ D {\bf 7}, 2317 (1973).
%%CITATION = PHRVA,D7,2317;%%
%41 citations counted in INSPIRE as of 24 Sep 2014
%
\bibitem{deser} 
S.~Deser,
%``Gravity from self-interaction redux,''
Gen.\ Rel.\ Grav.\  {\bf 42}, 641 (2010).
% [arXiv:0910.2975 [gr-qc]].
 %%CITATION = ARXIV:0910.2975;%%
 %8 citations counted in INSPIRE as of 24 Sep 2014
%
\bibitem{pw}
M.~K.~Parikh and F.~Wilczek,
%``Hawking radiation as tunneling,''
Phys.\ Rev.\ Lett.\  {\bf 85} (2000) 5042.
%[hep-th/9907001].
%%CITATION = HEP-TH/9907001;%%
%794 citations counted in INSPIRE as of 06 Jun 2014
%
\bibitem{davidson}
A.~Davidson and B.~Yellin,
%``Quantum Black Hole Wave Packet: Average Area Entropy and Temperature Dependent Width,''
Phys. Lett. B, 267 (2014).
%  [arXiv:1404.5729 [gr-qc]].
  %%CITATION = ARXIV:1404.5729;%%
  %1 citations counted in INSPIRE as of 15 Sep 2014%
%
\bibitem{ramy}
R.~Brustein,
%``Origin of the blackhole information paradox,''
Fortsch.\ Phys.\  {\bf 62} (2014) 255;
%[arXiv:1209.2686 [hep-th]];
%%CITATION = ARXIV:1209.2686;%%
%23 citations counted in INSPIRE as of 06 Jun 2014
R.~Brustein and M.~Hadad,
%``Wave function of the quantum black hole,''
Phys.\ Lett.\ B {\bf 718} (2012) 653.
%[arXiv:1202.5273 [hep-th]].
%%CITATION = ARXIV:1202.5273;%%
%6 citations counted in INSPIRE as of 06 Jun 2014
%
\bibitem{torres}
R.~Torres and F.~Fayos,
%``Singularity free gravitational collapse in an effective dynamical quantum spacetime,''
Phys.\ Lett.\ B {\bf 733} (2014) 169;
%[arXiv:1405.7922 [gr-qc]];
%%CITATION = ARXIV:1405.7922;%%
R.~Torres,
%``Singularity-free gravitational collapse and asymptotic safety,''
Phys.\ Lett.\ B {\bf 733} (2014) 21.
%[arXiv:1404.7655 [gr-qc]].
%%CITATION = ARXIV:1404.7655;%%
%1 citations counted in INSPIRE as of 06 Jun 2014
%
\bibitem{brustein}
R.~Brustein and A.~J.~M.~Medved,
%``Restoring predictability in semiclassical gravitational collapse,''
JHEP {\bf 1309} (2013) 015.
%[arXiv:1305.3139 [hep-th]].
%%CITATION = ARXIV:1305.3139;%%
%9 citations counted in INSPIRE as of 06 Jun 2014
%
\end{thebibliography}
\end{document}